\documentclass[a4paper,11pt]{article}
\pdfoutput=1 

\usepackage{jheppub} 

\usepackage[T1]{fontenc} 
\usepackage{amsmath}
\usepackage{graphicx}
\usepackage{array,multirow}
\usepackage{url}
\usepackage{fancyhdr}
\usepackage{lastpage}
\usepackage{tabularx}
\usepackage{rotating}
\usepackage[dvipsnames,table]{xcolor}
\usepackage{slashed}

\def\be{\begin{eqnarray}}
\def\ee{\end{eqnarray}}
\newcommand{\beq}{\begin{eqnarray}}
\newcommand{\eeq}{\end{eqnarray}}

\newcommand{\bs}[1]{{\boldsymbol #1}}
\def\nn{ \nonumber \\ }

\def\kap{{\bs \kappa}}
\def\kapone{{\bs \kappa}_1}
\def\kaptwo{{\bs \kappa}_2}
\def\dd{\text{d}}
\def\intq{\int_{\bs q}}
\def\q{{\bs q}}
\def\k{{\bs k}}
\def\p{{\bs p}}

\def\n{{\bs n}}
\def\x{{\bs x}}

\def\Q{{\tilde q }}

\def\pone{{\bs p}_1}

\def\epsk{\bs \varepsilon}
\def\Aone{\frac{\kapone-\q}{\left(\kapone-\q\right)^2}}

\def\Bone{\frac{\kapone}{\kapone^2}}

\def\Lip{{\bs L}}
\def\Avec{{\bs A}}
\def\Bvec{{\bs B}}
\def\Cvec{{\bs C}}
\def\tone{\tau_1}
\def\ttwo{\tau_2}
\def\tonetwo{\tau_{12}}
\def\W\mathcal{W}
\def\GamG{\Gamma_{\scriptscriptstyle G}}
\def\GamA{U_{\scriptscriptstyle G}}
\def\gamA{u_{\scriptscriptstyle G}}
\def\VG{ V_{\scriptscriptstyle G}}
\def\VA{U_{\scriptscriptstyle Q}}
\def\vA{u_{\scriptscriptstyle Q}}
\def\kh{k_{\scriptscriptstyle H}}

\def\kkh{\k_{\scriptscriptstyle H}}
\def\thh{\theta_{\scriptscriptstyle H}}
\def\ks{k_{\scriptscriptstyle S}}

\def\kks{\k_{\scriptscriptstyle S}}
\def\ths{\theta_{\scriptscriptstyle S}}
\def\xH{x^+_{\scriptscriptstyle H}}
\def\xS{x^+_{\scriptscriptstyle S}}
\def\kappaperpS{{{\bs \kappa}_{\scriptscriptstyle S}}}
\def\kappaS{\kappa_{\scriptscriptstyle S}}
\def\Ismall{{\scriptscriptstyle I}}
\def\Ssmall{{\scriptscriptstyle S}}
\def\Hsmall{{\scriptscriptstyle H}}
\def\Qsmall{{\scriptscriptstyle Q}}
\def\Gsmall{{\scriptscriptstyle G}}
\def\Dsmall{{\scriptscriptstyle D}}
\def\xs{{x^+}}

\def\QC{{\mathcal{Q }}}

\def\tauH{{\tau_{{\scriptscriptstyle H}}}}
\def\tauS{{\tau_{{\scriptscriptstyle S}}}}

\def\MqS{w_{\scriptscriptstyle Q}}
\def\MgS{w_{\scriptscriptstyle G}}
\def\Mantenna{\mathcal{P}^{(1)}_\text{ant}}
\def\Vscat{\mathcal{V}(\q)}

\def\kperpS{{\kks}}
 
\def\mkperpS{k_\Ssmall}
\def\mkperpH{k_\Hsmall}

\def\pplus{{p^+}}

\def\w{{k^+}}
\def\wH{{k^+_\Hsmall}}
\def\wS{{k^+_\Ssmall}}

\def\dd{{\rm d}}
\def\q{{\boldsymbol q}}
\def\mDebye{m_{\scriptstyle D}}

\def\A{\mathcal{A}}
\def\tauq{{ \frac{(\kperpS+\q)^2}{2 \wS}}}
\def\taug{{ \frac{(\kappaperpS+\q)^2}{2 \wS}}}
\def\tauqS{{\tau_{\scriptscriptstyle q}}}
\def\taugS{{\tau_{\scriptscriptstyle g}}}

\def\tauR{\tau_{\scriptscriptstyle \rm res}}
\def\rmed{{\lambda_{\scriptscriptstyle \rm res}}}
\def\thetaMed{\theta_{\scriptscriptstyle \rm med}}
\def\thetares{{\theta_{\scriptscriptstyle \rm res}}}

\def\Ab{{\bs A}_g} 
\def\L{{\bs L}_q}
\def\Aq{{\bs A}_q} 
\def\Bq{{\bs B}_q}
\def\Lb{{\bs L}_g}
\def\Bb{{\bs B}_g}

\def\W{\mathcal{C}}
\def\n{{\bf n}}
\def\thetaH{\thh}
\def\thetaS{{\ths}}
\def\nH{{ \n_\Hsmall}}
\def\nS{ {\n_\Ssmall}}

\newcommand{\Pel}[2] {\mathcal{P}^{({#1})}_{#2}}
\newcommand{\tfi}[2] {\tau^{({#1})}_{#2}}
\newcommand{\eq}[1] {eq.~(\ref{#1})}

\title{Jet formation and interference in a thin QCD medium}

\author[a,b]{Jorge Casalderrey-Solana,}
\author[a]{Daniel Pablos,}
\author[a,c]{Konrad Tywoniuk}

\affiliation[a]{
Departament d'Estructura i Constituents
de la Mat\`eria and Institut de Ci\`encies del Cosmos (ICCUB),
Universitat de Barcelona, Mart\'\i \ i Franqu\`es 1, 08028 Barcelona, Spain}
\affiliation[a]{
Rudolf Peierls Centre for Theoretical Physics, University of Oxford, 1 Keble Road, Oxford OX1 3NP, United Kingdom}
\affiliation[c]{
Physics Department, Theory
Unit, CERN, CH-1211 Gen\'eve 23, Switzerland}

\emailAdd{jorge.casalderreysolana@physics.ox.ac.uk}

\emailAdd{dpablos@ecm.ub.es}

\emailAdd{konrad.tywoniuk@cern.ch}

\abstract{
In heavy-ion collisions, an abundant production of high-energy QCD jets allows to study how these multiparticle sprays are modified as they pass through the quark-gluon plasma. In order to shed new light on this process, we compute the inclusive two-gluon rate off a hard quark propagating through a color deconfined medium at first order in medium opacity. We explicitly impose an energy ordering of the two emitted gluons, such that the ``hard'' gluon can be thought of as belonging to the jet substructure while the other is a ``soft'' emission (which can be collinear or medium-induced). Our analysis focusses on two specific limits that clarify the modification of the additional angle- and formation time-ordering of splittings. In one limit, the formation time of the ``hard'' gluon is short compared to the ``soft'' gluon formation time, leading to a probabilistic formula for production of and subsequent radiation off a quark-gluon antenna. In the other limit, the ordering of formation is reverted, 
which automatically leads to the fact that the jet substructure is resolved by the medium. We observe in this case a characteristic delay: the jet radiates as one color current (quark) up to the formation of the ``hard'' gluon, at which point we observe the onset of radiation of the new color current (gluon). Our computation supports a picture in which the in-medium jet dynamics are described as a collection of subsequent antennas which are resolved by the medium according to their transverse extent.
}

\begin{document} 
\maketitle
\flushbottom

\section{Introduction}
\label{sec:intro}

QCD jets are essential objects in modern particle physics. Many of the searches for new physics at the LHC involve the detailed analysis of the production and properties
of these energetic spays of particles that arise from the colour neutralisation of energetic partons produced in hadronic collisions. 
Jets also play a central role in the analysis of hot and dense matter formed in the debris of high energy Pb-Pb collisions (for a recent review of LHC heavy-ion results see \cite{Armesto:2015ioy}). Since long, these objects have been identified as the
 most powerful tomographic tools with which to diagnose the properties of the formed matter \cite{Bjorken:1982tu}. The tremendous  combined capabilities of the LHC (and its associated detectors) have converted this potential into reality; the copious production of jet samples at LHC energies enables detailed studies of jet properties in a heavy ion environment \cite{Aad:2010bu,Chatrchyan:2011sx,Chatrchyan:2012nia,Chatrchyan:2012gt,Chatrchyan:2012gw,Aad:2012vca,Aad:2013sla,Chatrchyan:2013kwa,Abelev:2013kqa,Chatrchyan:2013exa,Chatrchyan:2014ava,Aad:2014wha,Aad:2014bxa,Adam:2015ewa,CMSRAA,Adam:2015doa}. 
 
 Early LHC results on jet physics in Pb-Pb collisions at $\sqrt{s_{\scriptscriptstyle\text{NN}}}=2.76$ TeV have shown a strong suppression in the jet production rate as compared to proton-proton collisions at the same energies \cite{Aad:2014bxa,Adam:2015ewa,CMSRAA}. This reduction of the jet rate can be understood as a result of the energy loss experienced by jets on their way out of the collision zone. This phenomenon, known as jet quenching, was identified previously at RHIC by the observation of a strong suppression in the production of high energy hadrons in heavy ion collisions \cite{Adler:2003qi,Adams:2003kv}. While the latter suppression is mostly sensitive to the energy loss by the hardest jet fragments, the variety of the observed jet modifications in a heavy-ion environment demands addressing the jets as  sources of several partons propagating simultaneously through the QCD medium.
  
At high energies, parton energy loss is controlled by the stimulated radiation of medium-induced gluons as a result of the scattering with the medium constituents. Many of the properties of the modification of jets may be inferred from the single-gluon emission rate, first computed by BDMPS-Z \cite{Baier:1996kr,Zakharov:1996fv,Baier:1998kq}. In a finite length medium, an opacity expansion of the multiple scattering series resumed in BDMPS-Z was introduced in \cite{Wiedemann:2000za,Gyulassy:2000fs,Gyulassy:2000er}, in which the expansion parameter may be viewed as the ratio of the medium length to the mean free path.\footnote{This ratio is usually referred to as the medium opacity.} These computations are at the heart of the different formalisms later developed to address the dynamics of energetic partons in plasma \cite{Wang:2001ifa,Arnold:2002ja,Jeon:2003gi,Wicks:2005gt,Qin:2007rn,CasalderreySolana:2007pr,Majumder:2010qh,Ovanesyan:2011xy,Mehtar-Tani:2013pia,Qin:2015srf}. 
Generalising this picture in order to treat the interactions of QCD showers with the medium, one usually relies on working models that iterate the single-gluon emission rate without considering possible multi-parton correlations.
For different Monte-Carlo implementations, see \cite{Lokhtin:2008xi,Renk:2008pp,Armesto:2009fj,Schenke:2009gb,Zapp:2011ya}.
Nevertheless, in vacuum jet physics it has been long understood that interference effects between the shower constituents, known as coherent branching or angular ordering, are essential to completely describe intra-jet properties in high-energy colliders \cite{Ellis:1991qj}. 
 
For typical medium-induced gluons, all correlations are suppressed as their formation time over the medium length \cite{Blaizot:2012fh}, see also \cite{Apolinario:2014csa}. For large media, this allows to treat multiple medium-induced branchings in terms of a cascade \cite{Blaizot:2013hx,Kurkela:2014tla}, see also \cite{Baier:2000sb,Arnold:2002zm}. The large separation of scales, related in turn to the medium length and the mean free path, can also potentially lead to significant radiative corrections to transverse momentum broadening and energy loss in the medium \cite{Liou:2013qya,Blaizot:2014bha,Wu:2014nca} and, in general, to medium transport coefficients \cite{Blaizot:2014bha}. See also \cite{Ghiglieri:2015zma,Ghiglieri:2015ala} for related work in next-to-leading order corrections to the medium-induced spectrum. 
The emission spectrum of two gluons with comparable formation times was analysed for the case of a dense medium in \cite{Arnold:2015qya}.
These approaches, however, consider the gluon transverse momenta to be of the order of the medium scale and therefore do not explicitly study the interplay between emissions of the former kind and genuine vacuum emissions, either short- or long-distance ones, which can take place in the context of high-energy jets. In the present work, we aim at providing further analytical insight into these situations, complementary the numerical analysis of the rate at first order in medium opacity presented in \cite{Fickinger:2013xwa}.
 
The study of coherence effects for in-medium jets is a relatively new subject. In \cite{MehtarTani:2010ma,MehtarTani:2011tz,Armesto:2011ir,CasalderreySolana:2011rz,MehtarTani:2011gf,MehtarTani:2012cy} the single gluon emission rate off two classical colour currents was computed in different approximations. The main finding of these studies may be summarised as the emergence of a new scale, the medium resolution scale, which controls the ability of the medium to resolve the number of colour emitters that traverse the plasma. 
 If the transverse separation of the colour sources is larger than this scale, the medium is able to interact independently with each of the sources, and the medium-induced radiation spectrum consists of the superposition of the induced spectrum from the each of the colour currents. If the transverse separation is small compared to this scale,  the system interacts coherently with the propagating currents and the medium-induced spectrum coincides with that of a single colour charge in the overall colour representation of the system of currents. 
The phenomenon of medium resolution has also been recently found in the dynamics of energetic colour objects plunging through infinitely strongly coupled gauge theory plasmas \cite{Casalderrey-Solana:2015tas}. These findings lead to the suggestion of a new picture for jet dynamics in heavy-ion collisions where, from the point of view of the medium, the jet shower is organised in terms of effective emitters, according to the medium resolution scale \cite{CasalderreySolana:2012ef}. 
 
In this paper we study the coherent branching of soft gluons in jet showers by directly analysing the double-inclusive gluon emission rate. The two gluons have well separated energies, which are both much smaller than the quark energy. We will model the medium as a single scattering centre which interacts once with a jet shower at a given distance from the hard production vertex that generates the shower. This medium model corresponds to the leading order in opacity expansion for a medium which can be situated at any distance from the hard vertex. Since we can place the interaction at will, this scattering centre may be also viewed as a chronometer, which tests the jet shower at different times. The double inclusive emission rate in the $N=1$ opacity approximation was analysed in \cite{Fickinger:2013xwa} for realistic values of the medium parameters. However, our computation is not aimed at describing the gluon emission rate in a realistic model of the hot matter produced in ultra-relativistic heavy-ion collisions but rather to understand how in-medium interactions generically affect the gluon radiation pattern.



\section{Preliminaries}
\label{sec:Prelim}

We compute the inclusive rate off a hard quark that emits two soft gluons while it interacts with a single coloured scattering. In the absence of scattering centre, those gluons originate from the relaxation of virtuality of the microscopic process that generates the energetic quark. The presence 
of one scattering centre 
leads to a modification of the vacuum spectrum by changing the transverse momentum of one of the  gluons emitted at the production vertex.  In addition, the additional momentum transferred to the jet supplemented by the scattering centre leads to an additional source of radiation of gluons with transverse momentum of order of the momentum transfer. This process is the stimulated emission of gluons in the medium. We explicitly study the effect of interferences among these different physics processes in the final emission rate.
We will work in light-cone coordinates $X = (x^+,x^-,{\boldsymbol x} )$, where $x^\pm \equiv (x^0 \pm x^3)/2$ and $\x = \left(x_1,x_2 \right)$ denotes a transverse vector, $x \equiv |\x|$. For future reference, the momenta of the quark, the ``hard'' and the ``soft'' gluons are given by $P = (p^+, 0,{\boldsymbol 0})$, $K_{\scriptscriptstyle H} = (\kh^+, \kh^-, \kkh)$ and $K_{\scriptscriptstyle S} = (\ks^+, \ks^-, \kks)$, respectively. 

\subsection{Medium model}
\label{sec:MediumModel}

As the partons produced in a jet shower plough through a QCD medium, they exchange energy and momentum with its constituents. Since these interactions are mediated by the exchange of gluons, an effective way to encode those interactions is by analysing the propagation of energetic partons in a fluctuating colour gauge field, $A$, sourced by the quarks and gluons in the medium. 
For high-energy probes, the light-cone gauge $A^+ = 0$, with the plus-momentum component the largest momentum of the parton, is particularly convenient. In this gauge, typical fluctuating fields in the background will have all other components of comparable order. Since in the eikonal limit, which we will briefly describe in the next subsection, the coupling of a parton of momentum $P^\mu$ to the medium gauge field is proportional to $P \cdot A$, the contribution of the perpendicular field components $A^i$ to the probe-medium interactions is suppressed with respect to the contribution of the  $A^-$ component. 

The high-energy approximation also leads to simplifications in the momentum exchange with the medium. Assuming that the $p^+$ component of the probe is much larger than the momentum exchanged in the medium, $q$, the effect of the $q^+$  exchanged momentum is always suppressed with respect to the transverse momentum exchanges, $\q$, since by energy  momentum conservation in the vertex, the former is always added to the large parton momentum. This is equivalent to neglecting drag ($q^+ \approx 0$) for high-energy probes. With these assumptions, we can model the medium by a gauge  field with only one non-vanishing component, which takes the form
\beq
\label{eq:MediumPot}
A^{-}(Q) \equiv t^a A^{a,-}(Q) = 2\pi \delta(q^+) \, \int \dd x^+ e^{i q^- x^+} \, \A(x^+;\q) \,,
\eeq
where the medium field is real, $\A^\ast(x^+;\q) = \A(x^+;-\q)$.

The medium dynamics leads to the randomisation of the field. From the point of view of the probe, we may characterise the medium by cumulants of the fluctuating field configurations. Odd cumulants vanish as a consequence of colour neutrality of the medium. Even n-point medium correlators exhibit correlation lengths of order the inverse medium exchange. This allows us to approximate the medium average of the background gauge field to
\beq
\label{eq:MediumAverage}
\langle \A^{a}(x^+;\q) \A^{\ast\,b}(x'^+;\q') \rangle = \delta^{ab} \mDebye^2 n(x^+) \,\delta(x^+ - x'^+) \,(2\pi)^2 \delta(\q-\q') \Vscat \,,
\eeq
where $n(x^+)$ is the density of scattering centres in the $x^+$ direction. In the simplest of cases, a static medium with fixed density $n_0$ and length $L$, $n(x^+)=n_0 \Theta(L-x^+)$. As a consequence of Lorentz contraction, the correlation length along the $x^+$ direction can be neglected and we may consider exchanges as instantaneous in $x^+$. Furthermore, $\Vscat$ is the scattering potential, usually assumed to be  screened at the scale of the Debye mass $\mu_\Dsmall$. However, in our discussion the exact form of this potential does not matter as long as it is {\it isotropic} in the transverse plane.

The locality in $x^+$ of the correlator \eq{eq:MediumAverage} also implies that higher-order cumulants vanish in this high-energy approximation, such that higher-order medium correlators are simply products of the two point functions \eq{eq:MediumAverage}. The $N=1$ opacity approximation consists of describing all medium effects by the two point function \eq{eq:MediumAverage}. This is a good approximation when the medium is dilute, $\alpha_s n_0 L\ll 1$. If the medium is dense, a re-summation of an arbitrary number of medium exchanges is required. Nevertheless, in this paper we will employ the $N=1$ opacity approximation as a tool to explore the dynamics of energetic jet showers in QCD medium, but we will make no assumptions on whether this approximation correctly captures the properties of the quark-gluon plasma formed in heavy-ion collisions.

\subsection{Computing the amplitude and cross section}
\label{sec:Amplitude}

In our computation we assume the following energy ordering $p^+ \gg \kh^+ \gg \ks^+$. This is the conventional ordering of a vacuum shower leading to the double-logarithmic enhancement of gluon emissions. In the presence of the medium, such an ordering allows us to study how the hard jet substructure builds up in the presence of a probing, soft gluon which can be either collinear or medium-induced.

We also assume that the energy of both gluons is much larger than the typical momentum transfer from a medium exchange $q$, $\ks^+ \gg q$.  In order to consistently  neglect the radiation from the scattering centre, we will focus solely on collinear emissions with respect to the quark; this means that the emission angles of both of the gluons, $\thh = \kh/\kh^+$ and $\ths = \ks/\ks^+$,
are small. However, we will make no assumptions about the relative magnitude of the emission angles $\thh/\ths$. To account for all relevant medium effects, we will allow the transverse momentum of one of the gluons,  $\ks$, to be of the order of the in-medium momentum transfer $q$. 

These assumptions simplify significantly the computation of the emission amplitudes. Since we are focusing in the rate in which both the gluons and the quark posses a much larger energy than the medium momentum transfer, we can use an eikonal approximation for the QCD Feynman rules, which allows to exploit this separation of scales at amplitude level. 
For ease of computation we will work in the mixed representation where 
the minus-component of the momenta is Fourier transformed to configuration space. Although the eikonal Feynman rules are well-known, we have re-derived them in app.~\ref{sec:FeynmanDiag} to clearly state our approximations and for the readers convenience. To discuss how we have organised our computation, we proceed to list the main ingredients. 

In the eikonal limit, the quark and triple-gluon vertices conserve spin and helicity, respectively. It is useful to absorb the spin and polarization of the propagating degrees of freedom of adjacent propagators into properly contracted vertices, see app.~\ref{sec:FeynmanDiag}. In our present calculations, we will be using two types of eikonal emission vertices which are justified by the ordering of energies we have assumed. First, emissions of gluons off the energetic quark come with the factor
\beq
\label{eq:HardQuarkVertex}
\VG^{a,i}(p^+; \k_{\scriptscriptstyle I},k_{\scriptscriptstyle I}^+) \equiv \VG^{a,i}({\bs 0},p^+; \k_{\scriptscriptstyle I},k_{\scriptscriptstyle I}^+) = 2 ig \,t^a \frac{p^+}{k_{\scriptscriptstyle I}^+} \, k^i_{\scriptscriptstyle I}\,,
\eeq
where $I=H,S$, while the gluon splitting brings a factor. 
\beq
\label{eq:HardGluonVertex}
\GamG^{abc,i} (\kkh,\kh^+; \kks, \ks^+) = 2 g \, f^{abc} \frac{1}{z} \kappaS^i \,,
\eeq
where we have defined $z \equiv \wS/\wH$ and
\be
\label{eq:KappaDef}
\kappaperpS \equiv \kks - z\,  \kkh \,,
\ee
is the relative momentum of the emission. Both effective vertices are transverse vectors and proportional to the relevant colour factor. The four-gluon vertex does not exhibit the $1/z$ enhancement of \eq{eq:HardGluonVertex}, and it is therefore negligible in this kinematic limit. The same argument holds for the gluon splitting into a quark-antiquark pair.

The momentum in \eq{eq:KappaDef} deserves some discussion. Naively, the strong ordering in the energy of the gluon suggests that we may drop the apparently subleading contribution $z \kkh$ in the definition of $\kappaperpS$. However, this can only be done if $\kkh$  and $\kks$ are of the same order. This is the case, when both gluons are medium-induced. However, in this paper we will be interested in exploring the emission rate when the angle in transverse spacer $\thh$ and $\ths$ are comparable, which implies that $\kh$ is parametrically (in $1/z$) larger than $\ks$. 

Since there is only one non-vanishing component of the medium potential, the effective interaction vertices become scalars, and read
\begin{align}
\vA^a (p^+) &= 2 i g\, t^a \, p^+ \,, \\
\gamA^{abc} (p^+) &= 2 g\, f^{abc} \, p^+ \,,
\end{align}
where we have amputated the medium field in the definition of the vertex; for details see app.~\ref{sec:FeynmanDiag}.

Having absorbed all spinor and helicity structures into the vertices, the internal partons are simply described by scalar propagators.
Keeping the leading energy correction on the position of the poles, the 
propagation of the gluons in the mixed representation, see \eq{eq:ScalarPropagator}, may be expressed as 
\be
D_{\scriptscriptstyle G} (x^+ ; \k_{\scriptscriptstyle I}, k_{\scriptscriptstyle I}^+)\equiv D(x^+; \k_{\scriptscriptstyle I}, k_{\scriptscriptstyle I}^+) =\frac{\Theta(x^+)}{2 \w} \exp\left[-i \frac{\k_{\scriptscriptstyle I}^2}{2k_{\scriptscriptstyle I}^+}x^+ - \epsilon x^+ \right] \,,
\ee
where again $I=H,S$ and $\Theta(x)$ is the Heaviside theta-function. The propagation of the quark is identical to that of the gluon; however, since we are taking the quark to have a much larger energy, we will adopt the strict eikonal limit, $\kh/\pplus,\ks/\pplus, q/\pplus \ll1$  ,
\be
D_{\scriptscriptstyle Q}(x^+; \pplus ) \equiv D(x^+; {\bs 0}, p^+) = \frac{\Theta(x^+)}{2 \pplus}  e^{-\epsilon x^+}\,.
\ee
The $\epsilon$-prescription in the propagators above suppresses the propagation of modes in the distant past and future. Finally, all in-coming and out-going particles have to be multiplied by the appropriate phases, embodying energy-momentum conservation, and spinor or polarization vectors for quarks and gluon, respectively; see app.~\ref{sec:FeynmanDiag} for more details.


\begin{figure}[t!]
\centering
\includegraphics[width=0.7\textwidth]{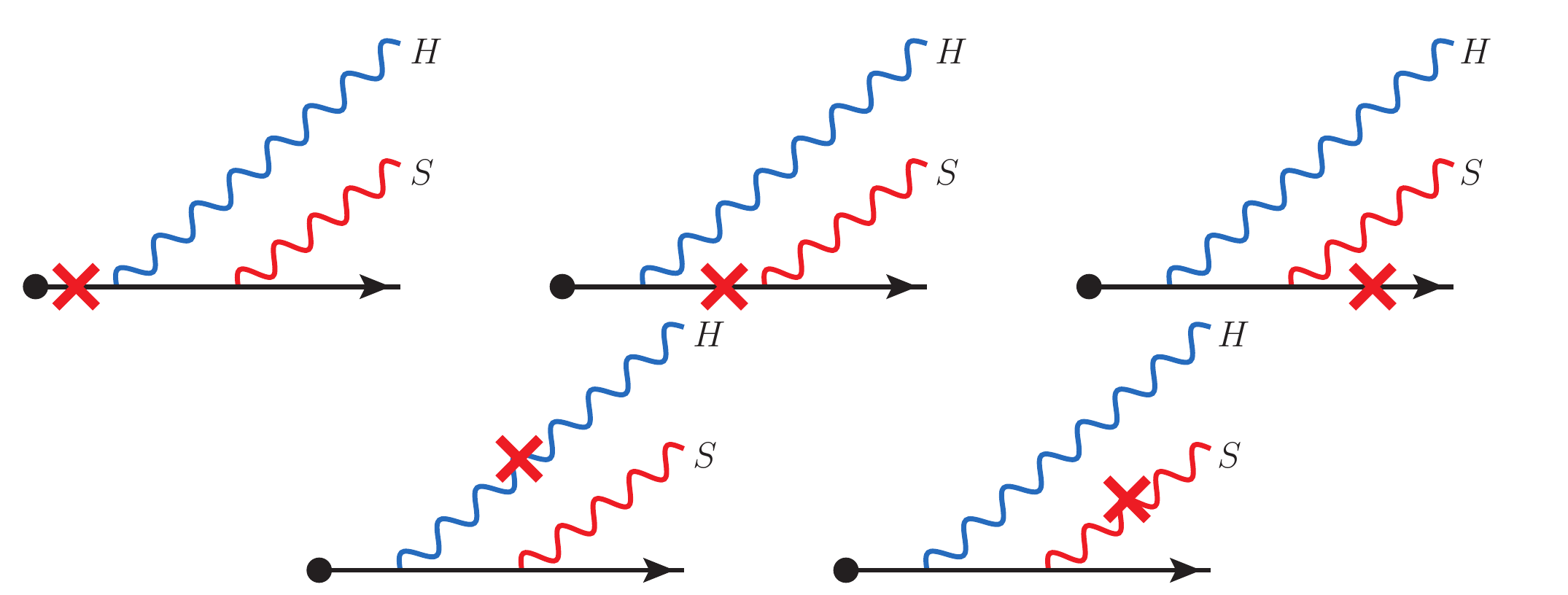}
\caption{Two-gluon emission off a quark with $N=1$ interactions with the medium. We also include the diagrams with the substitution $H \leftrightarrow S$.}
\label{fig:RealCombDiagrams} 
\end{figure}

\begin{figure}[t!]
\centering
\includegraphics[width=0.7\textwidth]{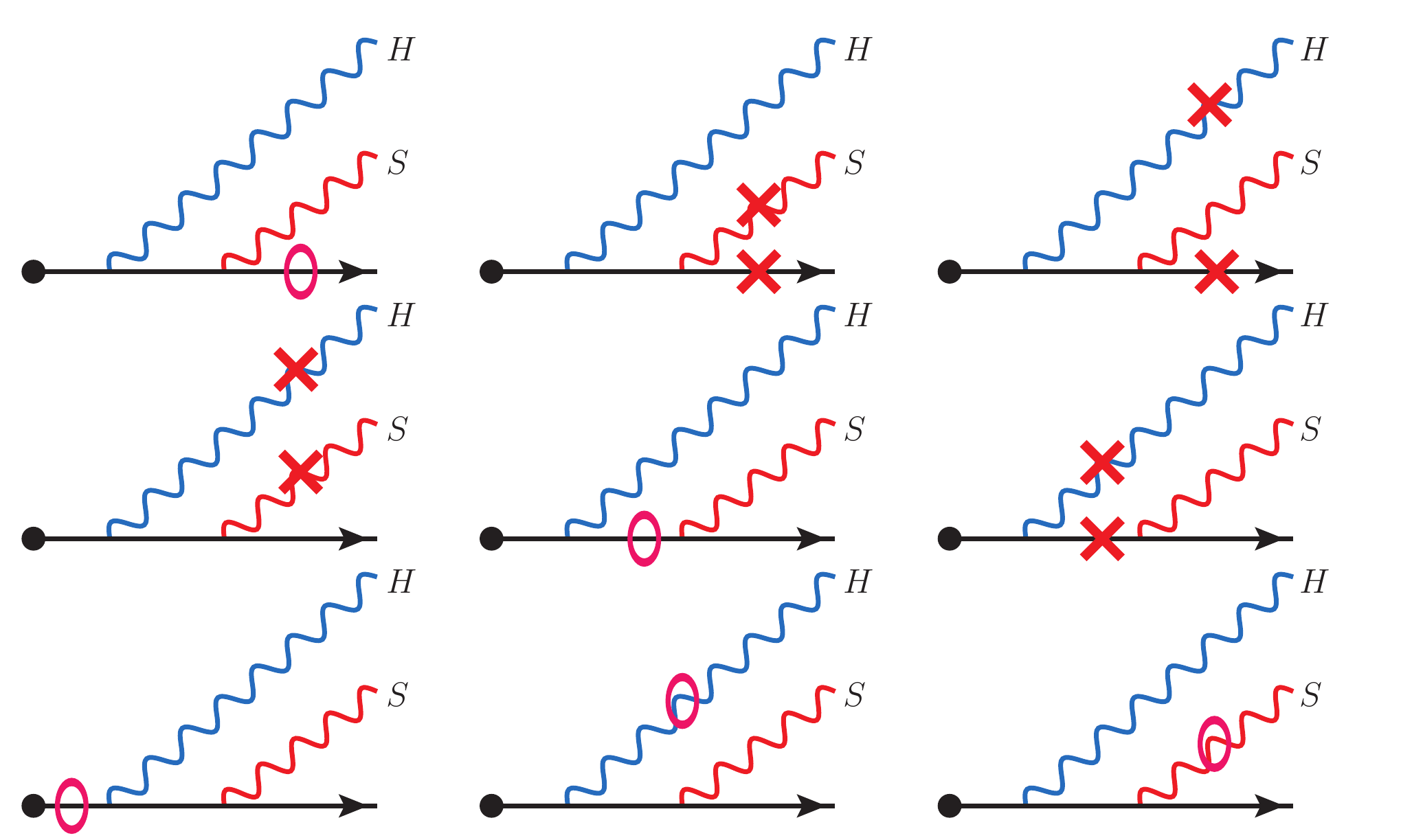}
\caption{Two-gluon emission off a quark with $N=1$ unitarity corrections. We also include the diagrams with the substitution $H \leftrightarrow S$.}
\label{fig:VirtualCombDiagrams} 
\end{figure}

\begin{figure}[t!]
\centering
\includegraphics[width=0.7\textwidth]{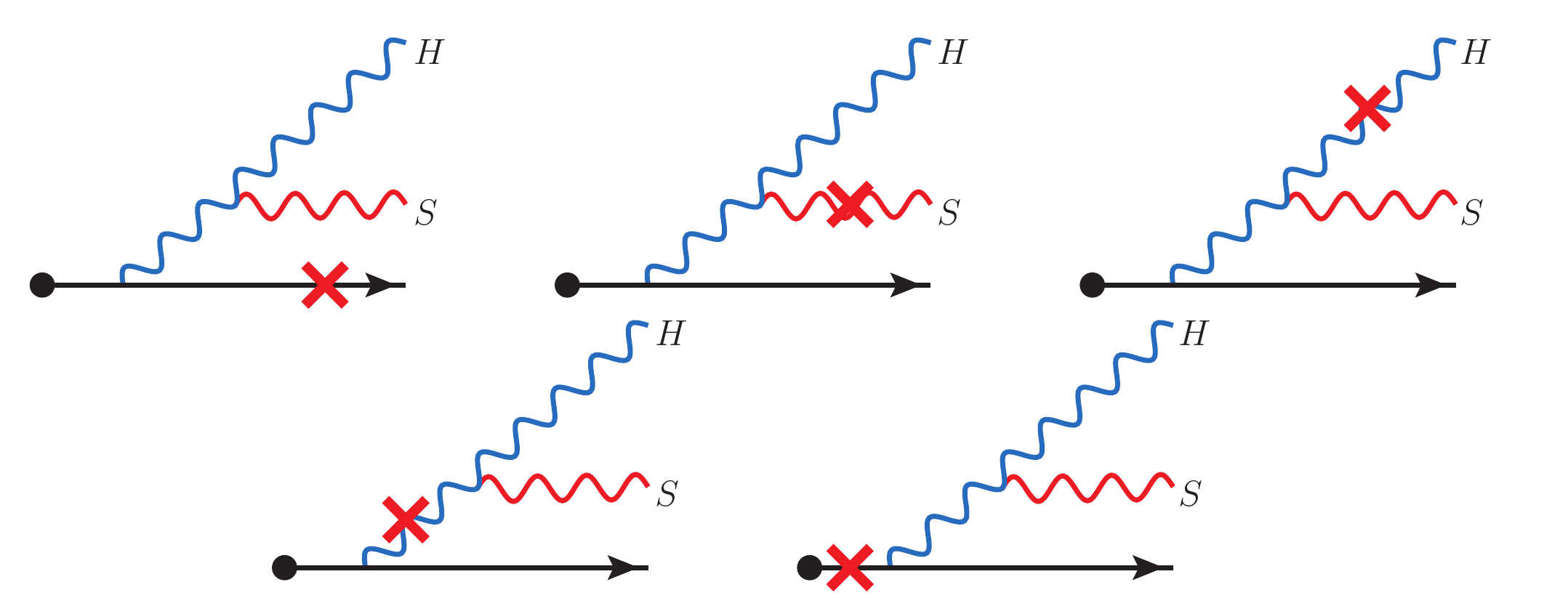}
\caption{Sequential two-gluon emission off the quark and  with $N=1$ interactions with the medium.}
\label{fig:RealForkDiagrams} 
\end{figure}

\begin{figure}[t!]
\centering
\includegraphics[width=0.7\textwidth]{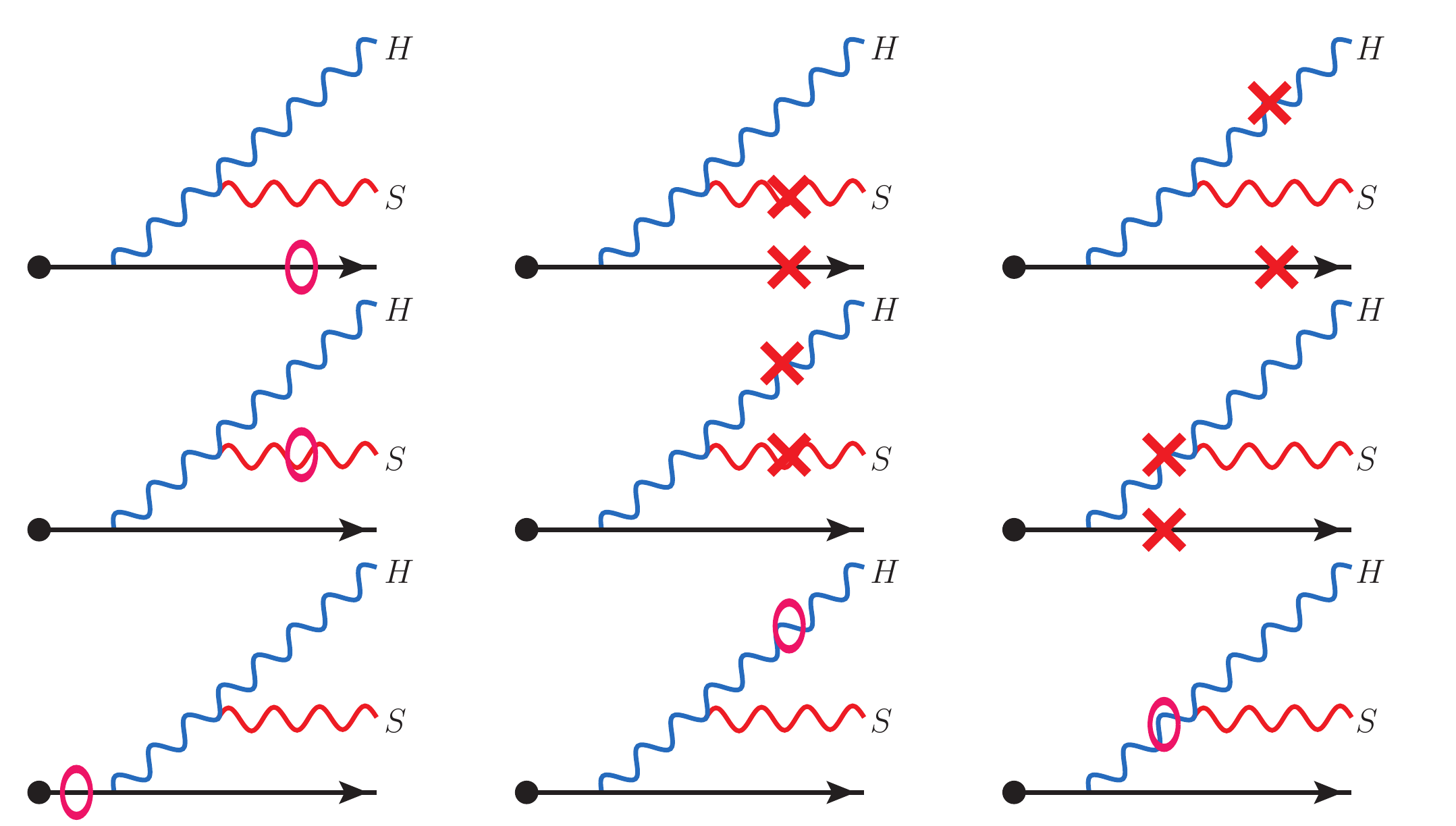}
\caption{Sequential two-gluon emission off a quark with $N=1$ unitarity corrections.}
\label{fig:VirtualForkDiagrams} 
\end{figure}

These eikonal rules are the building blocks with which we construct the double gluon emission rate. 
Neglecting the four-gluon vertex, the full computation of this process involves the calculation of 15 real amplitudes, summarised in Figs \ref{fig:RealCombDiagrams} and \ref{fig:RealForkDiagrams}. Also, a total 23 additional non-vanishing virtual corrections, or double-Born diagrams, which involve the interference between double scattering amplitudes with vacuum diagrams, need to be  considered. These are summarised in Figs. \ref{fig:VirtualCombDiagrams} and \ref{fig:VirtualForkDiagrams}. All relevant Feynman rules are summarised in app.~\ref{sec:ListFeynmanRules}. After squaring the real diagrams, the full rate is composed of 248 combinations.

Given the large number of diagrams that needs to be computed and squared, in this paper we have used an automated procedure to compute this cross section. We have coded the Feynman rules outlined in the previous section in Mathematica, and perform all integration, sums and colour algebra via symbolic computations. The amplitude of these processes can be written as
\begin{align}
\label{eq:AmpVacuum}
\mathcal{M}_{(0)\, \lambda \lambda'}^{ab} &= \varepsilon_\lambda^i(\kkh) \varepsilon_{\lambda'}^j(\kks) \, \sum_k m_{k}^{ab,ij} \,,\\
\label{eq:AmpMed1}
\mathcal{M}_{(1)\, \lambda \lambda'}^{ab} &= \int_{x^+}\int_{\q}\varepsilon_\lambda^i(\kkh) \varepsilon_{\lambda'}^j(\kks)\A^c(x^+;\q)  \sum_k m_{k}^{abc,ij}(x^+;\q) \,,\\
\label{eq:AmpMed2}
\mathcal{M}_{(2)\, \lambda \lambda'}^{ab} &= \int_{x^+.x'^+}\int_{\q,\q'} \varepsilon_\lambda^i(\kkh) \varepsilon_{\lambda'}^j(\kks) \A^c(x^+;\q) \A^d(x'^+;\q') \sum_k m_{k}^{abcd,ij} (x^+,x'^+;\q,\q')\,,
\end{align}
where the basic building blocks are the amputated amplitudes $m_{k}^{abXY,ij}(\ldots)$. The first two superscripts $\{a,b\}$ relate to the colour structure and the latter two $\{i,j\}$ relate to the indices of the two out-going transverse momenta and $\{\lambda, \lambda' \}$ are their respective polarizations. The superscripts $X$ and $Y$ and the number of arguments relate to the number of insertions of the medium field. In the argument of the function we have suppressed the kinematics of the emitted gluon. The subscript $k$ simply designates the particular diagram under consideration, and the sum runs over all diagrams in Figs.~\ref{fig:RealCombDiagrams} and \ref{fig:RealForkDiagrams} for \eq{eq:AmpMed1} and Figs.~\ref{fig:VirtualCombDiagrams} and \ref{fig:VirtualForkDiagrams} for \eq{eq:AmpMed2}. The vacuum terms are also trivially found from these diagrams. Thus \eq{eq:AmpVacuum} is the sum of all vacuum amplitudes, \eq{eq:AmpMed1} is the sum of diagrams with one medium insertion, and \eq{eq:AmpMed2} is the sum of all non-vanishing amplitudes with two medium insertions. 
We will refer to $\mathcal{M}_{(1)}$ as medium-real diagrams, and $\mathcal{M}_{(2)}$ will be referred to as medium-virtual diagrams.

The amplitudes in Eqs.~(\ref{eq:AmpVacuum})-(\ref{eq:AmpMed2}) are of course also all proportional to the amplitude of the hard process that created the out-going quark. Since this simply factorises into the Born cross-section for quark production, we will always suppress it.

To illustrate the procedure, we describe here how to compute one of the amputated amplitudes, namely the upper left diagram of  Fig.~\ref{fig:RealCombDiagrams}, which reads
\begin{align}
\label{eq:AmputatedAmplitude_Ex}
m_{1}^{abc,ij}(x^+;\q) &= \int_{\xH,\xS} e^{i\kh^- \xH + i\ks^- \xS} \, \VG^{a,i} (p^+;\kkh,\kh^+) \, D_{\scriptscriptstyle Q}(\xS-\xH;p^+) \nonumber \\
&\times \VG^{b,j} (p^+; \kks,\ks^+)\, D_{\scriptscriptstyle Q}(\xH-x^+;p^+) \,\vA^c(p^+)\,  D_{\scriptscriptstyle Q}(\xs-x_0^+;p^+) \,,
\end{align}
where the gluons are on-shell $\kh^- = \kkh^2/(2\kh^+)$ and $\ks^- = \kks^2/(2\ks^+)$. In order to alleviate the notation, we have also defined the integrals
\begin{align}
\int_{x^+} &= \int_{-\infty}^\infty \dd x^+ \,, \\
\int_{\q} &= \int \frac{\dd^2\q}{(2\pi)^2} \,,
\end{align}
and usually we will set $x_0^+=0$ if not stated otherwise.

In this amputated amplitude, $i$ and $j$ are transverse space indices that denote the transverse momentum component of the soft and hard gluons while $a$ and $b$ are their respective colour indices. The positions $\xH$, $\xS$ and $x^+$ denote the position in configuration space where the hard and soft emissions and the scattering with the medium take place. Finally, the two phase factors appearing in the integrand are a consequence of the external leg insertion. To deal with the colour algebra we use the ColorMath package \cite{Sjodahl:2012nk}.

Following these simple rules, we introduce all amplitudes in Mathematica, which we also use to square them. The $x_{\scriptscriptstyle H}^+$ and $x_{\scriptscriptstyle S}^+$ integrals in the amputated amplitude, see \eq{eq:AmputatedAmplitude_Ex}, and the transverse momentum multiplication are performed symbolically. Formally, we take advantage of the medium average \eq{eq:MediumAverage} to write for the square of the medium-real amplitudes
\be
\langle \left| \mathcal{M}_{(1)} \right|^2 \rangle &= \int_{x^+,x'^+} \int_{\q,\q'} \langle \A^{c}(x^+;\q) \A^{\ast\,c'}(x'^+;\q') \rangle \sum_{k,k'} m_{(k)}^{abc,ij}(x^+;\q)m_{(k')}^{\ast abc',ij}(x'^+;\q') \nn
&= \mDebye^2 \int_{x^+} \int_{\q} n(x^+) \Vscat\, \sum_{k,k'} m_{(k)}^{abc,ij}(x^+;\q)m_{(k')}^{\ast abc,ij}(x^+;\q) \,,
\ee
where the sum of all repeated indices is assumed and we have used $\sum_\lambda {\bs \varepsilon}_\lambda^i(\k) {\bs \varepsilon}_\lambda^j (\k)= \delta^{ij}$ in order to contract the transverse indices. A similar strategy can be followed for the medium-virtual amplitudes which are added by multiplying them with the vacuum amplitudes, such that
\begin{align}
\langle \mathcal{M}_{(2)}\mathcal{M}^\ast_{(0)} \rangle &=  \int_{x^+,x'^+} \int_{\q,\q'} \langle \A^{c}(x^+;\q) \A^{c'}(x'^+;\q') \rangle \sum_{k,k'} m_{(k)}^{abcc',ij}(x^+,x'^+;\q,\q') m_{(k')}^{\ast ab,ij}\nn
&= \mDebye^2 \int_{x^+} \int_{\q} n(x^+) \Vscat\, \sum_{k,k'} m_{(k)}^{abcc,ij}(x^+,x^+;\q,-\q)m_{(k')}^{\ast ab,ij}\,.
\end{align}
In particular, when calculating contact terms, i.e. diagrams with two insertions on the same propagator, denoted with a circle in Figs.~\ref{fig:VirtualCombDiagrams} and \ref{fig:VirtualForkDiagrams}, we use the half-value prescription for the Heaviside function, $\Theta(0)=1/2$, which yields the correct answer for the double-Born diagrams.
 The full medium-induced spectrum at first order in opacity is then
\begin{align}
\langle\left| \mathcal{M}_{\scriptscriptstyle \text{1OP}} \right|^2 \rangle &= \langle \left| \mathcal{M}_{(1)} \right|^2 \rangle + 2\text{Re} \langle \mathcal{M}_{(2)}\mathcal{M}^\ast_{(0)} \rangle \nn
&= \mDebye^2 \int_{x^+} \int_{\q} n(x^+) \Vscat\, w(x^+;\q) \,,
\end{align}
where
\begin{align}
w(x^+;\q) &= \sum_{k,k'} \Big[m_{(k)}^{abc,ij}(x^+;\q)m_{(k')}^{\ast abc,ij}(x^+;\q) + m_{(k)}^{abcc,ij}(x^+,x^+;\q,-\q)m_{(k')}^{\ast ab,ij} \nn
&\hspace{1em} + m_{(k)}^{ab,ij} m_{(k')}^{\ast\, abcc,ij}(x^+,x^+;\q,-\q) \Big] \,.
\end{align}
The spectrum of emitted gluons is then
\beq
\frac{\dd^2 N_{\scriptscriptstyle \text{1OP}}}{\dd \Omega_{\kh} \, \dd \Omega_{\ks}} \equiv \frac{1}{\sigma^\text{Born}_q} \frac{\dd^2 \sigma_{\scriptscriptstyle \text{1OP}}}{\dd \Omega_{\kh} \, \dd \Omega_{\ks}}= \frac{1}{2p^+} \langle\left| \mathcal{M}_{\scriptscriptstyle \text{1OP}} \right|^2\rangle \,,
\eeq
where the factor $1/(2p^+)$ is the quark flux and the phase space for the gluons is given by $\dd \Omega_k \equiv (2\pi)^{-3} \dd^2\k\, \dd k^+/(2k^+)$. Unitarity is enforced by demanding that the full $\langle\left| \mathcal{M}_{\scriptscriptstyle \text{1OP}} \right|^2\rangle \to 0$ when the medium momentum exchange vanishes, $q\to 0$. In other words, an exact cancellation of medium-real and medium-virtual diagrams takes place in this limit, such that no double-counting with the pure vacuum cross section is allowed. We have explicitly checked that our expressions respect this condition.

\section{Analysis of the induced rate}
\label{sec:AnalysisRate}

The strategy of computing the full amplitude in an automatised form allows us to deal with the many diagrams we have computed in a simple and effective way. However, the answer that this computation yields is lengthy and we have not been able to reduce it to a simple form. Therefore, in this section we will explore two particularly interesting limits of this 
expressions in which we have managed to express the answer in a closed form. This analysis is complementary to the numerical analysis of the full rate performed in \cite{Fickinger:2013xwa}.

Prior to taking these two limits, some general considerations about the full square matrix are in order. 
The analysis of the colour structure of the both the real and virtual contributions allows us to separate the full answer into only two non-vanishing 
colour structures, after averaging over colours. The total rate may be expressed as
\be
\label{eq:cs_general}
w(x^+;\q) = C_F^2 C_A \MqS (x^+;\q)+ C_F C^2_A \MgS(x^+;\q) \,,
\ee
where the elements $ \MqS$ and $\MgS$ are functions of the kinematic variables.
This general structure admits a simple interpretation. The full rate may be understood as the sum of two different physical processes: i) the emission of two gluons off the high-energy quark, $\MqS$; and  ii) the emission of a hard gluon off the high-energy quark which, in turn, emits an additional soft gluon $\MgS$. These emissions can occur either as a result of the initial virtuality of the hard vertex that creates the energetic quark or as a result of the interaction of the system with the medium. Note that in the infinite quark energy limit, the change of momentum of the quark as a result of the interaction and emission processes is negligible. This implies that, effectively, only the gluon scatters and terms proportional to $C_F^3$ are suppressed by powers of the quark energy.

The expressions for $\MqS$ and $\MgS$ may also be organised according to their dependence on the position of the scattering centre, $\xs$.
In general, we can express these two terms as
\be
\label{eq:gen_decomp}
w_\Ismall(x^+;\q) =\sum_{i=1}^{N_\Ismall} \Pel{i}{\Ismall}(\q) \Big\{1-  \cos \big[ \xs / \tfi{i}{\Ismall}(\q) \big] \Big\}
\ee
where $I=Q,\,G$, $N_\Qsmall=2 $ and $N_\Gsmall=19$ is the number of independent terms for the two distinct colour structures and the functions $ \Pel{i}{\Qsmall}$, $ \Pel{i}{\Gsmall}$, $\tfi{i}{\Qsmall}$, $\tfi{i}{\Gsmall}$ are rational functions of products of the four momenta of the three partons as well as of the transferred momenta. 
These are, in general, complicated expressions which we have not been able to simplify to a compact form, and therefore we shall not present them here. Instead, in the following subsections, we will show the results of these in two distinct kinematic limits.

It is interesting to note that all the dependence of the full rate,  \eq{eq:gen_decomp}, 
 on the position of the scattering centre, $\xs$, occurs in the form of cosine-like phase factors. 
These phases indicate interference effects between the vacuum production of the quark, at $x_0^+=0$, and the medium scattering processes.
For single gluon emission these interferences are well known, and are the precursors of the Landau-Pomeranchuk-Migdal (LPM) effect. In the context of radiation induced by a single scattering, this effect can be understood as the frustration of the induced radiation when the scattering centre is placed too close to the hard vertex, at a distance shorter than the formation time of the emitted gluon, $\tau= 2 k^+/(\k- \q)^2$. While in the single emission rate all the cross section is characterised by this single time scale, in the double-gluon emission rate several time scales appear.

This distinct $\xs$ dependence of the total rate allows us to treat differently the kinematic factors $\Pel{i}{\Ismall}$ and the formation factors $\tfi{i}{\Ismall}$ when expanding the rate in different kinematic regions. The limits we will explore invoke certain assumptions about the momenta of the partons and the transferred momenta, for $any$ medium length $L$. In fact, the dependence of any scale related to the medium length only enters in the computation via the phase factors, as $L$ appears in the limits of integration of $\xs$. Therefore, interferences are solely dependent on the relative magnitude of the formation factor $\tfi{i}{\Ismall}$ and the medium length. For this reason, when taking  kinematic limits, it is consistent to expand the kinematic factors $\Pel{i}{\Ismall}$ and $\tfi{i}{\Ismall}$ to different orders since apparently subleading terms in $\tfi{i}{\Ismall}$ may be enhanced by the medium length. In the next two subsections we will specify the limits we explore and describe this approximation in more detail.

\subsection{Expansion parameters}
\label{subsec:EP}

We now specify the parameters which we use to expand the symbolically computed cross section. First of all, as it is clear from the Feynman rules in sec.~\ref{sec:Amplitude}, the energy of the quark disappears form the final rate, since there is a cancelation between the $\pplus$-dependence of the eikonal vertices and the eikonal quark propagators. This is only true in the strictly infinite quark energy limit which we adopt. Secondly, the structure of these rules also indicates that the rate depends on the
energies of both the emitted gluons via the combination 
\be
z=\frac{\ks^+}{\kh^+} \, ,
\ee
which, by assumption, is small.

We analyse the double emission rate as a function of the emission angles of the two gluons. Assuming both these angles are small, these are trivially related to the momentum of the emitted gluons via 
\be
\theta_\Hsmall = \frac{\kh}{\kh^+} \,, \quad \quad \theta_\Ssmall = \frac{\ks}{\ks^+} \,.
\ee
In terms of these angles, the variable $\kappaperpS$, defined in \eq{eq:KappaDef}, is independent of $z$.
To make this scaling explicit,  we may write $\kappaperpS $ as
\be
\label{kappa_def_angles}
\kappaperpS=\wS \left( \thetaS  \, \nS-  \thetaH \, \nH\right)
\ee
 with $\n_\Ssmall$ ($\n_\Hsmall$) the unit transverse vector along the direction of $\kks$ ($\kkh$).
This form motivates us to organise the computation in terms of the (dimensionless) ratio of angles 
\be
\label{eq:rDef}
r = \frac{\theta_\Hsmall}{\theta_\Ssmall}  \,.
\ee
In addition to the momenta of both gluons, the medium interaction introduces an additional dimension-full quantity, namely the momentum transferred by the scattering centre, $\q$. In order to properly take limits of the full rate, we need to consider the relative magnitude of this momentum transfer to other dimension-full quantities in the rate. Motivated by the fact that the single-gluon medium-induced rate is dominated by gluons with transverse momenta of order the transferred momentum, we choose to organise our computation in terms of the (dimensionless) ratio
\be
\label{eq:qscaling}
\Q=\frac{q}{\ks}= \frac{1}{z \theta_\Ssmall } \frac{q}{ \wH} \,.
\ee
This ratio  ensures that, as long as we keep $\Q$ finite, the softest gluon in the amplitude may be medium-induced. This choice of scaling introduces a non-trivial dependence of the emission rate on the variable $z$. This may be best illustrated by considering the ratio between the transferred momentum and the transverse momentum of the hard gluon
\be
\label{eq:qKHratio}
\frac{q}{\mkperpH}= \Q \, \frac{z}{r} \,.
\ee
Although by construction we have assumed that $z$ is small in the Feynman rules, the introduction of the scaling \eq{eq:qscaling} leads to a different behavior depending on the relative magnitude of $z$ and $r$. By keeping $\Q$ fixed,  the limit  $z\ll r$ implies that  the transverse momentum of the hard gluon is much greater than the medium momentum transfer; complementary, the limit in which $r\ll z$, the transverse momentum of the hard gluon is much smaller than the momentum transfer. These are the two limits that we will explore in the next subsection. We will leave the analysis of the region  $r\sim z$, in which the transverse momentum of both gluons are comparable to the medium transfer, for future work.

These two particular limits also have a close relation to two distinct space-time pictures of the emissions. In the limit of small $z$, the hard gluon is formed early in the medium. The two hardest partons, the quark and the hard gluon, will therefore form an effective dipole, or antenna, that is probed by the emission of the softest parton in the cascade. This situation is close in spirit to the one studied in \cite{MehtarTani:2010ma,MehtarTani:2011tz,Armesto:2011ir,CasalderreySolana:2011rz,MehtarTani:2011gf,MehtarTani:2012cy}. In the opposite limit, the formation times of the soft gluon is shorter than the hard one, allowing it to be emitted earlier in the cascade. This is a novel situation that we study for the first time in detail. 

\subsection{Emission rate in the soft limit}
\label{sec:EarlyAntenna}

We start by studying the double induced gluon rate in the limit $z\ll r$, i.~e.,
\be
\label{eq:Limit1}
z\rightarrow 0\,, \quad {\rm with} \quad \left\{r,\, \Q,\, \theta_S, \wH \right\} \, \,\rm{fixed}  \,.
\ee
As already discussed, in this limit \eq{eq:qKHratio} implies that $\mkperpH$ decouples from the medium scale. This implies that the medium-induced rate of the hard gluon is power suppressed, and, to leading order in $z$, the rate of emission of this gluon is dominated by the vacuum processes associated to the hard vertex. Note also that the (vacuum) formation times of the two gluons also are strongly ordered, 
\be
\label{eq:tauRL1}
\frac{\tauH}{\tauS}=\frac{z}{r^2} \,,
\ee
with $\tauH= 2 \wH /\kkh^2$ and $\tauS=2 \wS/\kks^2$ the vacuum formation times of the hard and soft gluons, respectively. Therefore, at leading order in $z$, the hard gluon is effectively emitted from the hard vertex since it is formed at an arbitrarily short time. 

We expand the prefactors $\Pel{i}{\Qsmall}$ and $\Pel{i}{\Gsmall}$ in \eq{eq:gen_decomp} to leading order in $z$, which means that we only keep term which diverge as $z^2$. In this limit, only 1 out of the 2  $\Pel{i}{\Qsmall}$ and 10 out of the total 19 $\Pel{i}{\Gsmall}$ terms remain. The phase factors of those terms  $\tfi{i}{\Qsmall}$ and  $\tfi{i}{\Gsmall}$, posses different leading $z$ behavior, $z^{-1}$.   
Following the discussion around \eq{eq:gen_decomp} (on the presence of a new scale, the position of the scattering centre from the hard vertex, $\xs$) induces us to expand the phase factors 
$\tfi{i}{\Qsmall}$, $\tfi{i}{\Gsmall}$ to a different $z$-order than the prefactors $\Pel{i}{\Qsmall}$ and $\Pel{i}{\Qsmall}$. Performing this limit, several cancellations occur and we ultimately were able to bring the cross section to a closed form.


Following the notation of \eq{eq:cs_general}, the term associated to the emission of the two gluons by the quark, proportional to the colour factor $C_F^2 C_A$, term in the leading z-limit is given by,
\be
\label{eq:MQL1}
\MqS (x^+;\q) = \frac{4g^2}{\kkh^2}\, \times \, \left(-8 g^4\right)\frac{\kks \cdot \q}{(\kks+\q)^2 \, \kks^2} \left\{ 1- \cos\left[ \tauq \xs\right]\right\}\,,
\ee
where we have expressed all the factors in terms of products of the final momenta. This expression admits a simple physical interpretation. This part of the cross section is the product of the vacuum probability of emitting a gluon off the hard quark (in the soft limit), $\sim 1/\kkh^2$, times the $N=1$ opacity spectrum for the medium-induced emission of a soft gluon by the same quark  \cite{Gyulassy:2000fs, Wiedemann:2000za}. For later convenience, let us define the vectors,
\be
\label{eq:ALdefs}
\Aq =\frac{\kks+\q}{\left(\kks+\q\right)^2} \,, \quad  \Bq = \frac{\kks}{\kks^2} \,, \quad  \L = \Aq -\Bq \,,
\ee
where the latter vector is often referred to as the Lipatov vertex. In terms of these functions, the $N=1$ opacity expression for the single gluon emission may be expressed as  \cite{Gyulassy:2000fs, Wiedemann:2000za}
\be
\label{eq:N1asymptotic}
\frac{-\kks\cdot \q }{\kks^2 (\kks+\q)^2} = \frac{1}{2} \left( \L^2 + \Aq^2 -\Bq^2 \right) \,.
\ee
We will refer back to this decomposition in sec.~\ref{conclusions}.

We now turn to the $C_F C_A^2$ contribution in the same limit. After expanding the prefactors and the phase factors in the manner described above, we find convenient to express the full answer in terms of the vectors
\be
\label{eq:ALbdefs}
\Ab = \frac{\kappaperpS+\q}{\left(\kappaperpS+\q\right)^2}  \, , \quad  \Bb = \frac{\kappaperpS}{\kappaperpS^2} \, ,  \quad  \Lb= \Ab -\Bb \,,
\ee
where $\kappaperpS$ has been defined in \eq{eq:KappaDef}. These functions are analogous to those of \eq{eq:N1asymptotic} but, since they are functions of $\kappaperpS$ they may be viewed as the transverse momentum of the soft gluon as measured with respect from the hard one. In terms of these functions, the contribution to the cross section in which the hard gluon emits the soft one is given by 
\begin{align}
\label{eq:MGL1}
\MgS(x^+;\q) & = \frac{4g^2}{\kkh^2} \times 4 g^4 \left\{ \left(\Lb^2+\Ab^2-\Bb^2 - \Aq \cdot \Lb \right)  \left\{ 1- \cos\left[\taug \xs \right] \right\} \right. \nn 
&\hspace{2em} -\L \cdot \Ab \left\{ 1-\cos\left[\tauq \xs \right]\right\} \nn
&\hspace{2em}+\L \cdot \Lb \left\{ 1-\cos \left[ \left(\taug-\tauq\right) \xs \right]\right\} \nn
&\hspace{2em} \left. +\W \left(\wH,\kkh; \wS, \kks\right) \sin \left[\frac{\kks^2}{2 \wS} \xs \right] \sin\left[\frac{\q \cdot \kkh}{\wH} \xs\right] \right\} \,,
\end{align}
where we have defined the function\footnote{In spite of the explicit fraction  $\wS/\wH$ in \eq{eq:Wfunc}, this term is of the same $z$-order as the rest of the terms in \eq{eq:MGL1}.}
\be
\label{eq:Wfunc}
\W\left(\wH,\kkh; \wS,\kks \right) =- \frac{1}{4} \frac{\wS}{\wH} \frac{\kappaperpS \cdot \kkh}{\kkh^2\, \kks^2\, \kappaperpS^2}
\ee
Since $\W$ does not depend on $\q$, after integration over transferred momenta with the isotropic distribution $\Vscat$ in \eq{eq:MediumAverage}, the last term of \eq{eq:MGL1} vanishes. Therefore, similarly to \eq{eq:MGL1},  this contribution to the full rate is also proportional to the vacuum probability of emitting a hard gluon, $\sim 1/\kkh^2$. Combining \eq{eq:MQL1} and \eq{eq:MGL1} we may express the full answer for the double-inclusive gluon rate as 
\be
\label{eq:EAfac}
\left. \langle \left|\mathcal{M}_\text{1OP}\right|^2 \rangle \right|_{z\ll r} = \mathcal{P}_{\rm vac} \left(\mkperpH\right) \times \Mantenna \left(\ks \right) \,,
\ee
where $\mathcal{P}_{\rm vac}$ is the vacuum $q\to q+g$ splitting function in the soft limit 
\be
\label{def:Pvac}
\mathcal{P}_{\rm vac}(\kh) = \frac{2 C_F\, g^2}{\mkperpH^2} \,, 
\ee
and $\Mantenna$ is the emission rate of soft gluons off a hard quark-gluon dipole at first order in medium opacity, which we have derived in app.~\ref{sec:AntennaDerivation} using the method of classical currents employed in \cite{MehtarTani:2011gf}.  $\Mantenna$ is given in \eq{eq:Ampl2Medium_1} and \eq{eq:Ampl2Medium_Color}, which we reproduce here for the readers convenience,
\begin{align}
\label{antenna_spectrum}
\Mantenna  & =  4 g^4 C_A \mDebye^2 \int_{x^+}\int_{\q} n(x^+) \Vscat \nn
&\times \Bigg\{ C_F (\L^2 + \Aq^2 - \Bq^2) \left\{ 1- \cos \left[ \tauq \xs \right] \right\} \nn
&\hspace{1em} +C_A \Bigg[ \left(\Lb^2+\Ab^2-\Bb^2 - \Aq \cdot \Lb \right)  \left\{ 1- \cos\left[\taug \xs \right] \right\} \nn 
&\hspace{2em} -\L \cdot \Ab \left\{ 1-\cos\left[\tauq \xs \right]\right\} \nn
&  \hspace{2em}+\L \cdot \Lb \left\{ 1-\cos \left[ \left(\taug-\tauq\right) \xs \right]\right\} 
\Bigg] \Bigg\} \,,
\end{align}
which only depends on two dynamical time scales, 
\be
\label{def_tauS}
\tauqS = \frac{2\ks^+}{(\kks+\q)^2} \, , \quad \quad \taugS=\frac{2\ks^+}{(\kappaperpS+\q)^2} \, 
\ee
which are the formation times of the soft gluon when emitted collinear to the hard quark, $\tauqS$, and to the hard gluon, $\taugS$.

The factorised form \eq{eq:EAfac} admits a simple physical interpretation of the emission rate. In this limit, in which the hard gluon cannot be medium-induced, its production is totally dominated by vacuum physics, and its rate is determined by the vacuum splitting function. Since, as shown in \eq{eq:tauRL1}, the formation time of the hard gluon is parametrically smaller than that of the soft gluon, the hard gluon is emitted early. From the point of view of the medium, the system behaves as a quark-gluon antenna right after the hard vertex. Therefore, the scattering centre interacts with this two-parton systems simultaneously. The emission pattern includes several interferences effects, encoded in the intricate $\xs$ dependence of the rate, which are a result of the simultaneous propagation of this multi-parton system. We will discuss this pattern in detail in sec.~\ref{conclusions}.

\subsection{Emission rate in the collinear limit}
\label{sec:LateAntenna}

We now explore a different limit for the emission rate in which the hard gluon is not forced to be emitted first. To do so, we will consider the emission rate in the limit in which  the ratio of angles goes to zero first ($r \rightarrow 0$) and  then look for terms leading in the  ratio of energies $z \ll 1$. The scaling of the momentum transferred with the transverse momentum of the medium gluon, \eq{eq:qscaling}, implies that the order of limits does not commute, since, as expressed in \eq{eq:qKHratio}, in this limit the transverse momentum of the hard gluon, $\mkperpH$, becomes parametrically smaller than $q$.
The limit we performed may be summarised as,
\be
\label{eq:def_L2}
r\rightarrow 0,\, z\rightarrow 0, \quad {\rm with} \quad \left\{ \Q,\, \theta_S, \wH \right\} \, \,\rm{fixed}  \,.
\ee 
 It is easy to see from \eq{eq:tauRL1} that in this limit the formation time of the hard gluon is parametrically longer than that of the soft emission.
  
As before,  we expand the pre-factors $\Pel{i}{\Qsmall}$ and $\Pel{i}{\Gsmall}$ to leading order in $r$, which  corresponds to $r^{-2}$. 
In this limit, only 1 out of the 2 terms in $\Pel{i}{\Qsmall}$ and 7 out of the 19 terms $\Pel{i}{\Gsmall}$ are non-vanishing, and all of them 
depend on $z$ as a power, $z^{-2}$.  
Unlike the previous case,  not all the phase factors of the surviving terms possess the same leading $r$ limit: 6 of the phase factors $\tfi{i}{\Gsmall}$ are $\mathcal{O}(r^0)$ and the remaining one is $\mathcal{O}(r^2)$. The presence of the additional scale $\xs$ allows us to keep the latter apparently suppressed time scale, which becomes important at times of order $r^{-2}$. Nevertheless, in the leading $r$ limit, all 6 terms with $\mathcal{O}(r^0$) phases cancel identically, leading to 
\begin{align}
\label{eq:MQL2}
\MqS (x^+;\q) &= \frac{4 g^2}{\kkh^2}\, \times\, (-8g^4) \frac{\kks \cdot \q}{\kks^2\, (\kks+\q)^2} \left\{1- \cos\left[\tauq \xs\right]\right\} \,,
\\
\label{eq:MGL2}
 \MgS (x^+;\q) &= \frac{4g^2}{\kkh^2}\, \times \,4g^2\frac{q^2}{\kks^2\,\left(\kks+\q\right)^2 } \left\{ 1-\cos \left[\frac{\kkh^2}{2 \, \wH} \, \xs\right] \right\} \,.
\end{align}
As in the previous limit, both terms of the emission amplitude are proportional to $\mathcal{P}_{\rm vac}$, \eq{def:Pvac}, which implies that the production of the hard gluon proceeds as in vacuum. This is a consequence of the fact that in this limit $\mkperpH$ is parametrically smaller than $q$. Since as a consequence of the LPM interference, the medium-induced rate is collinear finite \cite{Gyulassy:2000fs, Wiedemann:2000za},  gluons with $\mkperpH\ll q$ cannot be medium-induced, and, therefore, they can only originate in the hard vertex that creates the jet.

The soft gluon emission rate depends on two distinct time scales. The emission rate of soft gluons off  the quark, 
$\MqS (x^+;\q)$, is identical in both limits, \eq{eq:MQL2} and \eq{eq:MQL2}, which, as we have discussed coincides with the $N=1$ opacity spectrum. This spectrum is controlled by the formation time of the soft gluon in medium $\tauqS$.
 The contribution to the rate coming from emissions off the hard gluon, 
$\MgS (x^+;\q)$, is different in this limit. This rate no longer depends on the time scale $\taugS$, as \eq{eq:MGL1}. The time scale controlling interferences in \eq{eq:MGL2} is the formation time of the hard gluon, $\tau_\Hsmall = 2 \wH/\kkh^2$. 
 
The appearance of $\tau_H$ in the emission rate leads to a simple consequence. If the scattering centre interacts with the quark-gluon system prior to the formation of the hard gluon, $\xs\ll \tauH$, the emission rate is dominated by radiation off the quark, since the \eq{eq:MGL2} vanishes. After $\tauH$, the emission rate may be understood as the incoherent superposition of the radiation off the quark plus the radiation off the gluon. At these late times, the rate of emission off the hard gluon is given by Gunion-Bertsch term \cite{Gunion:1981qs}, 
 \be
 \label{def:gb}
\lim _{r\rightarrow 0 } \Lb^2=\frac{\q^2}{\kperpS^2\,(\kperpS+\q)^2 } \, ,
 \ee
where $\Lb$ is defined in \eq{eq:ALbdefs}, which corresponds to the emission of a soft gluon of momentum $\kperpS$ off a hard on-shell gluon generated infinitely far away from the scattering centre. This rate differs from the emission rate of the soft gluon off the quark, \eq{eq:MQL2}, which is the $N=1$ spectrum. 

This difference in the emission rate  implies that after formation of the gluon, this new source of colour does not radiate in-medium as an independent new source produced at $\tauH$, which would lead to an equivalent $N=1$ spectrum off the hard gluon. The origin of the different rates at asymptotic late times may be understood from the analysis of the antenna spectrum, 
\eq{antenna_spectrum}. First of all, we note that in the limit \eq{eq:def_L2}, the time scales that control the emission from the colour dipole are parametrically suppressed with respect to $\tauH$. In the small r, small limit
\be
\label{antennascalesvanish}
\frac{\tauH}{\tauqS}= \mathcal{O} \left(\frac{z}{r^2}\right) \, ,\quad \frac{\tauH}{\taugS} = \mathcal{O}  \left(\frac{z}{r^2}\right)\, ,\quad \frac{\tauH}{\tauqS}-\frac{\tauH}{\taugS}=\mathcal{O}  \left(\frac{z}{r}\right) \,.
\ee
Because of this separation of time-scale, the relevant limit of the antenna spectrum 
\eq{antenna_spectrum}
is to consider $\xs\rightarrow \infty$, which implies that all phase factors average out to zero. In this incoherent limit, together with \eq{eq:def_L2} , the part of the antenna spectrum proportional to $C_A$ is 
\be
\lim_{r\rightarrow 0}
\left(
\Lb^2 + \left(\Ab^2 - \Aq \cdot \Ab \right)-  \left(\Bb^2 - \Bq \cdot \Bb \right)
\right)=
\frac{\q^2}{\kperpS^2\, (\kperpS+\q)^2 } \, ,
\ee
which coincides with the Gunion-Bertsch rate found in the two-gluon cross section.

This observation leads to a simple interpretation of the full double emission rate. If the scattering centre is placed early compare to $\tauH$, the interaction of the vacuum jet with the medium is identical to the interaction of a hard quark with the scattering centre; the emission of the hard gluon occurs after the scattering, and proceeds as in vacuum. If, on the contrary, the interaction with the scattering centre occurs at a time long compared to $\tauH$, then the hard gluon has time to form via vacuum processes which leads to the generation of an in-medium antenna.  From this time on, it is the quark-gluon dipole the one that interacts with the medium. We will discuss the consequences of this interpretation in sec.~\ref{conclusions}.

\section{Discussion and Outlook}
\label{conclusions}

In this paper we have discussed two particular kinematic regions, summarised in \eq{eq:Limit1} and \eq{eq:def_L2}, of the emission rate of two gluons in a thin (opacity $N=1$) medium. These limits are particularly interesting because they allow us to cleanly separate vacuum and medium emissions. As we have discussed, in both those kinematic regions, the emission of one of the gluons, the hard gluon, is dominated by vacuum-like processes produced at the hard vertex that creates a jet, while the soft gluon is medium-induced. By choosing this kinematics, we have focused on understanding how the multi-parton state associated to the propagation of a jet in plasma interacts with a QCD medium.  However, this limit prevents us from studying how the emission of new partons by medium-induced processes interferes with  the evolution of the jet shower, which demands the analysis of the rate when both emitted gluons can be medium-induced. This is a more intricate analysis which we leave for future work. Other analyses of the rate of emission of two gluons of comparable momentum from an on-shell quark propagating in the plasma can be found in \cite{Fickinger:2013xwa,Arnold:2015qya}.

As we have shown, within these regions the emission pattern or soft gluons is controlled by the emission spectrum off a hard quark-gluon dipole in medium, \eq{antenna_spectrum}. This type of dipoles has been recently used to understand emissions by  multiple colour sources in plasma and led to a rich interference structure \cite{MehtarTani:2010ma,MehtarTani:2011tz,Armesto:2011ir,CasalderreySolana:2011rz,MehtarTani:2011gf,MehtarTani:2012cy}. The particular case of opacity $N=1$ for a colour singlet antenna was analysed in detail in \cite{MehtarTani:2011gf}. Although the main lessons of the interference emission pattern of the quark-gluon antenna may be inferred form the analysis of the colour singlet antenna in \cite{MehtarTani:2011gf}, for completeness we discuss those features below.

The antenna emission rate, is controlled by three distinct time scales, the in medium formation times of soft gluons emitted off the quark, $\tauqS$, and off the gluon, $\taugS$, defined in \eq{def_tauS}
and a  third time scale,  intrinsically multi-partonic, which combines kinematic information of both constituents of the antenna,
\be
\label{tauR_def}
\tauR^{-1} =\frac{1}{\tauqS } - \frac{1}{\taugS} =\frac{2 \q -\kks-\kappaperpS}{2} \, \n
\ee
with $\n=\kkh/\wH$ a vector in transverse space whose modulus is the opening angle of the quark gluon system, $\n^2=\thetaH^2$. This time-scale controls the interference between medium-induced emissions of the two sources.  

To understand how these interferences occur, let us first consider the induced  spectrum off the quark-gulon dipole when the angle of the antenna $\theta_H$ is large. 
In this limit, the stimulated emissions off the quark and off the hard gluon are independent of one another. 
 The medium-induced spectrum off each of the propagating sources is dominated by gluons emitted with a typical transverse momentum {\it with respect to the source} of order $\mDebye$; this means that medium-induced gluons off the quark have $\mkperpS\sim\mDebye$ while the induced gluons off the hard gluon have $\kappaS\sim \mDebye$. Therefore, the induced spectrums will be well separated from one another if the angle of the dipole  $\thetaH\gg \thetaMed$, with $\thetaMed=\mDebye/\wS$, the typical emission angle with respect to the emitting source. In this limit, as a consequence of \eq{kappa_def_angles}, medium-induced gluons off the quark have $\kappaS\approx  \wS \thetaH\gg \mkperpS$ and induced gluons off the hard gluon have $\mkperpS \approx \wS \thetaH\gg \kappaS$.
 This condition is sufficient to show that, up to corrections of order $\thetaMed^2/\thetaH^2$, the antenna spectrum \eq{antenna_spectrum}, is the incoherent superposition of the stimulated spectrum off the quark  and off the hard gluon. 

Complementary, soft gluon emission off the antenna suffer strong interferences for  $\thetaH\ll \thetaMed$. In this limit, 
for typical induced gluons
 $\kappaperpS \approx \kperpS \sim \mDebye$, which implies that $\tauqS\approx\taugS$,   $\Aq\approx\Ab$, $\Bq\approx \Bb$ and $\L\approx \Lb$. The antenna spectrum is reduced to
 \be
 \label{small_angle_antenna}
\left.  w^{(1)}_\text{ant}\left(x^+; \kks,\ks^+\right)\right |_{\thetaH\ll \thetaMed} &= &C_F \left[1- \cos \frac{\xs}{\tauqS} \right] \left(\L^2 + \Aq^2 -\Bq^2 \right) \nn
 & +&C_A \left[1-\cos\frac{\xs}{\tauR} \right] \L^2 \,,
 \ee
 which is the medium-induced spectrum off the quark plus and additional interference term encoding the emission off the gluon. 
 In the small dipole limit, for typical induced gluons  \eq{tauR_def} yields $\tauR^{-1}\sim \mDebye\thetaH$. Therefore, interferences between the two sources suppress the emission off the hard gluon if at the relevant observation time $\xs$ the transverse size of the dipole, $\lambda=\thetaH \xs \ll \rmed$, with
 \be
 \rmed=\frac{1}{\mDebye} \,,
 \ee 
 the {\it transverse resolution scale}, which in this dilute medium equals the inverse typical momentum transfer by the medium.
 
 The characteristic time-scale for medium-induced radiation is the formation time of the emitted gluon, $\tauqS$. At this time, the typical transverse size of the quark gluon dipole is $\lambda \sim \rmed \, \thetaH/\thetares$. Therefore, for induced gluons with $\thetares\gg \thetaH$ the transverse size of the dipole at emission is small compared to the transverse resolution scale and the spectrum is totally dominated by the emission off the hard quark. Nevertheless, 
 the condition $\thetares\gg \thetaH$ depends on the frequency of the soft gluon, and at fixed $\thetaH$ only a fraction of the induced spectrum, with $\wS\ll \mDebye/\thetaH$,  is suppressed by 
 interference effects.  
   Since LPM interference suppresses induced radiation with formation time larger than the medium length $L$, the medium-induced spectrum has a  maximum frequency of emission, $\omega_{max}\sim \mDebye^2 L$ \cite{Wiedemann:2000za,Gyulassy:2000fs,Gyulassy:2000er}.  Therefore, if the dipole opening angle $\thetaH \ll 1/\mDebye L$, 
all the medium-induced spectrum off the gluon is cancelled by interference, 
  and the full emission spectrum off the quark gluon dipole is given by the medium-induced radiation off a hard quark. The multipartonic system interacts with the plasma as a single colour charge, the total charge of the system,  as long as
  its maximal transverse size in the medium  $\thetaH L\ll \rmed$.  
As we have seen, the rich interference structure associated to the medium resolution scale $\rmed$
 emerges in the soft limit of the double gluon emission rate, \eq{eq:EAfac} in which the emission of a hard gluon may be viewed as the production of an in-medium antenna.

The emergence of the antenna interference pattern  suggests a simple organising principle to understand the dynamics of jet showers in medium based on the resolution scale, as already suggested in \cite{CasalderreySolana:2012ef}. 
A basic element of this picture is that, similarly to vacuum,  jet showers may be best understood as a collection of in-medium antennas, that are dynamically generated in the process of relaxation of the virtuality of the jet. Our computations in the small angle regime support this picture. As we have seen in sec.~\ref{sec:LateAntenna}, the emission of soft gluons by a quark-gluon system generated in vacuum explicitly depends on the formation time of the gluon $\tauH$. While for short times $\xs\ll \tauH$ the emission pattern off the quark-gluon system is just that off a quark, after $\tauH$, both the quark and the hard gluon  contribute to the emission spectrum. This agrees with the common approximation of considering $\tauH$ as the time in which the hard gluon decorrelates from the hard quark. Nevertheless, disregarding the trivial differences in colour factors, the emission spectrums of the quark and gluon are not identical. Therefore, a naive  iteration of the medium-induced spectrum after the formation of the hard gluon leads to the incorrect emission rate.

 As we have shown, the emission rate off the hard gluon is predicted from the antenna picture in the small angle limit, provided the antenna forms after $\tauH$. The physical origin of the discrepancy between the iteration of the $N=1$ opacity spectrum, \eq{eq:MQL2},  and the antenna prediction is easy to understand. As it is well known, in the totally incoherent limit ($\tauqS\ll \xs$) the emission rate can be expressed as \eq{eq:N1asymptotic}, which we reproduce hare for convenience,
 \be
 \frac{-2 \, \kperpS \cdot \q}{(\kperpS+\q)^2 \kperpS^2} =\L^2+ \Aq^2 -\Bq^2\,, 
 \ee
which shows that the emission rate is the sum of  the stimulated emission off and on-shell quark, $\L^2$, the medium broadening of a soft gluon produced in the hard vertex, $\Aq^2$, and a unitarity correction which subtracts strength from the vacuum emission rate of soft gluons  off the quark, $\Bq^2$. The difference in the in-medium rate off the hard gluon lies precisely in a different $vacuum$ emission rate  of soft gluons by the quark-gluon antenna. As shown in \eq{eq:Vacuumqqg}, because of the interference with the quark, the vacuum emission of soft gluons off the hard gluon is proportional to
\be
\mathcal{P}^\Gsmall_{\rm vac}\propto\left( \Bb-\Bq\right)^2\,,
\ee
where the dependence in $\Bq$ reflects the interference with the quark. This vacuum rate vanishes in the limit of $\thetaH\ll\thetaS$, since, on the formation time of the vacuum soft gluon, the transverse size of the quark-gluon dipole is negligible. This does not imply that the in-medium rate vanishes, since, in the limit \eq{eq:def_L2}, $\tauH\sim 1/\wH \thetaS r^2$ and the dipole has a (parametrically)
 long time to separate, leading to a large transverse separation at formation time $\Delta=\thetaH \tauH \sim  z/r \mkperpS$.  As a result of vacuum interference, the emission off the gluon coincides with the emission off an on-shell gluon generated infinitely far away from the scattering centre, since no soft gluons are produced in vacuum in this kinematic lint. 
  
The small angle approximation explored in sec.~\ref{sec:LateAntenna} is, however, insensitive to the the medium resolution scale $\rmed$. 
 For typical in medium radiation with $\ks \sim \mDebye$, at the time the hard gluon forms, the traverse size of the dipole $\lambda = \rmed \,z/r \gg \rmed$ and the antenna is totally resolved. For this reason, at times long compared to $\tauH$ the two gluon emission rate, \eq{eq:MQL2} and \eq{eq:MGL2}, coincides with the spectrum of a fully resolved ($\xs \rightarrow 0$) small angle antenna, \eq{small_angle_antenna}. Unfortunately, the limit in which $r\rightarrow 0$ first is inadequate to explore the interplay between $\rmed$ and the formation of the hard gluon. In future work we plan to address the limit $r\sim z\rightarrow 0$ to understand how the formation of the antenna affects the resolution of the colour structure of the propagating dipole.

\acknowledgments

We are grateful to  Nestor Armesto, Edmond Iancu, Yacine Mehtar-Tani, Carlos Salgado and Derek Teaney
for  helpful discussions. DP acknowledges the hospitality of the Center for Theoretical Physics, MIT, where part of this work was completed.
JCS is a University Research Fellow of the Royal Society.
The work of JCS was  also supported by a Ram\'on~y~Cajal fellowship.  The work of JCS and DP was 
supported by  the  Marie Curie Career Integration Grant FP7-PEOPLE-2012-GIG-333786, by grants FPA2013-46570 and  FPA2013-40360-ERC
and MDM-2014-0369 of ICCUB (Unidad de Excelencia `Mar\'ia de Maeztu') 
 from the Spanish MINECO,  by grant 2014-SGR-104 from the 
Generalitat de Catalunya 
and by the Consolider CPAN project.

\appendix

\section{Effective Feynman Rules in the eikonal limit}
\label{sec:FeynmanDiag}

Here we will shortly review the relevant Feynman diagram technique applied in the light-cone gauge in the mixed representation, which is quite similar to the so-called time-ordered perturbation theory.
Choosing the gauge vector to be purely ``minus'', $n \equiv (0,1,{\bs 0})$, leads to the gauge condition $n\cdot A = A^+ = 0$, for the gluon field. In particular, the gluon polarization vector becomes
\beq
\varepsilon_\lambda (k) = \left(0, \frac{\k \cdot \epsk_\lambda(\k)}{k^+}, \epsk_\lambda(\k) \right) \,.
\eeq

%

\subsection{Propagators}

The scalar (Feynman) propagator for massless particles in vacuum reads 
\beq
D (k) = \frac{i}{k^2 + i\epsilon} \,,
\eeq
In terms of this, the quark and gluon propagators read
\begin{align}
S (k) &= \sum_s u^s(k){\bar u}^s(k) \, D (k) \,,\\
G ^{\mu\nu}(k) &= \sum_\lambda \varepsilon_\lambda^{\ast\,\mu}(k) \varepsilon_\lambda^\nu(k) \, D (k) \,,
\end{align}
respectively, where
\begin{align}
\sum_s u^s(k){\bar u}^s(k) &= \slashed{k} \,,\\
\label{eq:PolarizationSum}
\sum_\lambda \varepsilon^{\ast \,\mu}(k) \varepsilon^\nu(k) &= - g^{\mu\nu} + \frac{k^\mu n^\nu + k^\nu n^\mu}{k\cdot n} - k^2 \frac{n^\mu n^\nu}{\left(k\cdot n \right)^2} \,.
\end{align}
The transverse part of the polarization vectors satisfy $\sum_\lambda \epsk^{\ast\,i}_\lambda \epsk^j_\lambda= \delta^{ij}$.
Since all propagating partons are put on-shell, the latter term in \eq{eq:PolarizationSum} is irrelevant for our analysis.
The gluon propagator is symmetric, and the only non-vanishing components read
\beq
G ^{--}(k) = \frac{\k^2}{\left(k^+ \right)^2} D (k) \,, \hspace{1em} G ^{-i}(k) = \frac{k^i}{k^+} D (k) \,, \hspace{1em} G ^{ij}(k) = \delta^{ij} D (k) \,,
\eeq
where $i=(1,2)$. In the next subsection we will show that in the eikonal limit all the vertices become transverse and diagonal in spin and polarization, which allows them to absorb all the quark/gluon dependence (numerators). This leaves us with the momentum flow and pole structure which are encoded exlusively in the scalar part of the propagator.

We will work in the mixed representation, with (light-cone) time, energy and transverse momentum. For instance, energy-momentum conservation in a $q(l) \to q(p) + g(k)$  takes the following form
\beq
\label{eq:EMConservation}
(2\pi)^4\delta^{(4)}(p+k-l) = (2\pi)^3\int \dd x^+ \, e^{i \left(p^- + k^- - l^- \right)x^+} \, \delta\left( p^+ + k^+ - l^+  \right) \delta\left({\bs p} + \k - {\bs l} \right) \,,
\eeq
and similarly for the four-gluon vertex. 
While external particles are naturally demanded to be on-shell, this also allows to put all internal propagators on-shell, by
\begin{align}
\label{eq:ScalarPropagator}
D(x^+;k)\equiv D(x^+; \k,k^+ ) &= \int_{-\infty}^{\infty} \frac{\dd k^-}{2\pi} e^{-i k^-x^+} \, D(k) \,,\\
&= \frac{\Theta\left( x^+\right)}{2 k^+} e^{- i k^-x^+ - \epsilon x^+} \,,
\end{align}
with $k^- = \k^2/(2k^+)$ in the last line.
The $\epsilon$-prescription regulates the behavior of the propagator at infinity. We note that we have to keep in mind that any integration of an {\it external} point, involves the additional phases as in \eq{eq:EMConservation}.

\subsection{Leading eikonal vertices}

Here we derive the relevant vertices and demonstrate their behavior in the eikonal approximation. That means, we only keep terms that are enhanced by a factor $z^{-1}$, where $z$ is the fraction of a small energy over a large one. This automatically leads to the preservation of (quark) helicity and (gluon) polarization that floats through the vertices. 

The triple gluon vertex reads
\beq
i V_{abc}^{\mu\nu\sigma}(K_1,K_2,K_3) = g f^{abc} \big[(K_1-K_2)^\sigma g^{\mu\nu} + (K_2-K_3)^\mu g^{\nu\sigma} + (K_3-K_1)^\nu g^{\mu\sigma} \big] \,,
\eeq
where all momenta are incoming and $f^{abc}$ is the SU(3) structure constant. Enforcing energy-momentum conservation in the vertex, we will define the emission vertex describing $g(K_1+K_2) \to g(K_1) + g(K_2)$, where the momentum flow follows the time flow, as
\begin{align}
\label{eq:TripleGluonVertex}
&\GamG^{abc}(K_1,K_2) = \varepsilon^\ast_\mu(K_1) \varepsilon^\ast_\sigma(K_2) \, i V_{abc}^{\mu\nu\sigma}(-K_1,K_1+K_2,-K_2) \,\varepsilon_\nu(K_1+K_2) \nn
&\hspace{1em} = 2 g f^{abc} \left[\frac{1}{z} \left(\kap \cdot \epsk_2 \right)\left(\epsk_2 \cdot \epsk_{12} \right) + \left(\kap \cdot \epsk_1 \right)\left(\epsk_2 \cdot \epsk_{12} \right) -\frac{1}{1+z} \left(\kap \cdot \epsk_{12} \right)\left(\epsk_1\cdot \epsk_{2} \right) \right] \,,
\end{align}
where $\kap \equiv \k_2 - z \k_1$, $z\equiv k_2^+/k_1^+$ and ${\bs \varepsilon}_i \equiv  {\bs\varepsilon}(\k_i)$. In the eikonal approximation we only keep the leading $z$ term in \eq{eq:TripleGluonVertex}. We will however keep the apparently sub-leading contribution to $\kap$ since $|\kap| \sim k^+(\theta_1 + \theta_2)$ and we are interested in arbitrary angular ordering. The eikonal triple-gluon vertex becomes completely transverse and reads
\beq
\label{eq:EmissionVertexGluon}
\GamG^{abc,k}(\k_1,k_1^+;\k_2,k_2^+) = 2g f^{abc}\, \frac{1}{z}\, \kappa^k \,,
\eeq
where we have dropped a diagonal matrix for the propagation of the polarization.
This ensures that the polarization of the hardest gluon is conserved in the vertex.
On the other hand, including one gluon field from the medium in the triple gluon vertex leads to
\begin{align}
\label{eq:3gVertex}
&\varepsilon^\ast_\mu(K_1) A^c_\sigma(K_2) \, i V_{abc}^{\mu\nu\sigma}(-K_1,K_1-K_2,K_2) \,\varepsilon_\nu(K_1-K_2) \nn
&\hspace{15em} = 2  g f^{abc} k_1^+  A^{c-}(K_2)\, \epsk_1\cdot \epsk_{12} \,.
\end{align}
Using the decomposition in \eq{eq:MediumPot} for the medium potential, and applying the transformation to the mixed representation allows to define a triple-gluon interaction vertex
\begin{align}
\label{eq:InteractionVertexGluon}
\GamA^{ab}(x^+;k^+,\q) &= 2g\,k^+ \, f^{abc} \A^c(x^+,\q) \,, \\
&= u_{\scriptscriptstyle G}^{abc}(k^+) \, \A^c(x^+,\q) \,.
\end{align}
where the transversality of the vertex again is suppressed.

Following the same approach as for the gluons, we will define properly contracted QCD vertices that absorb the numerators of the propagators. For the emission of a gluon, $q(K_1+K_2) \to q(K_1)+g(K_2)$, we define
\beq
\label{eq:EmissionVertexQuark1}
\VG (K_1,K_2)= {\bar u}^{t}(K_1)\big(ig \gamma^\mu t^a \big) \varepsilon^{\ast\,\mu}_\lambda(K_2) \,u^s(K_1+K_2) \,,
\eeq
where $t^a$ is the SU(3) generator in the fundamental representation. In the eikonal limit, we take advantage of ${\bar u}^t(K_1) \gamma^\mu u^s(K_1+K_2) = 2 K_1^\mu \delta^{ts} + \mathcal{O}(K_2/K_1)$ which conserves spin, to define
\beq
\label{eq:EmissionVertexQuark2}
\VG^{a,i}(\k_1,k_1^+;\k_2,k_2^+) = 2ig \,t^a\, \frac{1}{z} \, \kappa^i \,,
\eeq
where again $\kap \equiv \k_2 - z \k_1$ and $z \equiv k_2^+/k_2^+$ and we have suppressed a diagonal matrix for the spin components. For the interaction with the medium we simply replace the polarization vector $\varepsilon^{\ast\,\mu}_\lambda(K_2)$ in \eq{eq:EmissionVertexQuark1} by the medium field $A^\mu(Q)$, and find
\begin{align}
\label{eq:InteractionVertexQuark}
\VA(x^+;k^+,\q) &= 2ig \,k^+ \,t^a \A^a(x^+,\q) \,, \\
&= u_{\scriptscriptstyle Q}^a(k^+) \,\A^a(x^+,\q) \,.
\end{align}
We note that the newly defined emission vertices, \eq{eq:EmissionVertexGluon} and \eq{eq:EmissionVertexQuark2}, allow us to absorb all spin and polarization information contained in the numerator of the quark and gluon propagators. We are therefore left with scalar propagators and the vertices for emission and interaction. The interaction vertices, \eq{eq:InteractionVertexGluon} and \eq{eq:InteractionVertexQuark}, are scalars as well.

\subsection{Effective Feynman rules}
\label{sec:ListFeynmanRules}

As a summary, we provide a list of graphical rules that can be used in order to calculate any diagram in light-cone perturbation theory in the mixed representation.
\begin{align}
\begin{minipage}{5cm} \includegraphics[width=\textwidth]{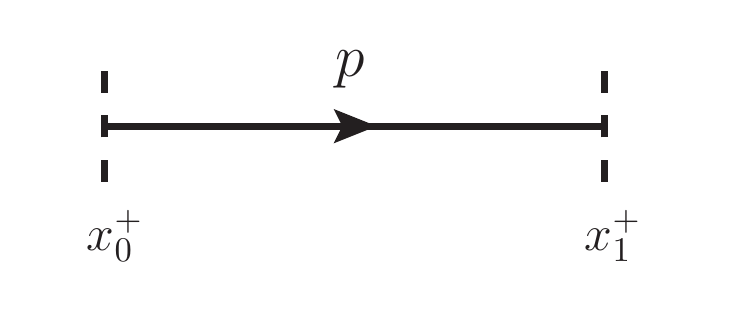} \end{minipage} &= \hspace{0.5em} D\left(x_1^+ - x_0^+; p^+ ,\p\right) \,,\\
\begin{minipage}{5cm} \includegraphics[width=\textwidth]{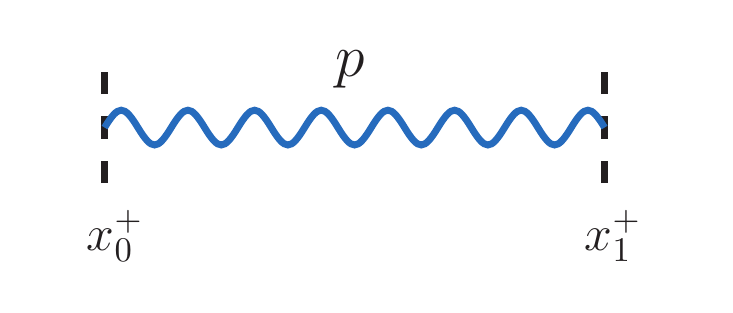} \end{minipage} &= \hspace{0.5em} D\left(x_1^+ - x_0^+; p^+ ,\p\right) \,,
\end{align}
\begin{align}
\begin{minipage}{4.5cm} \includegraphics[width=\textwidth]{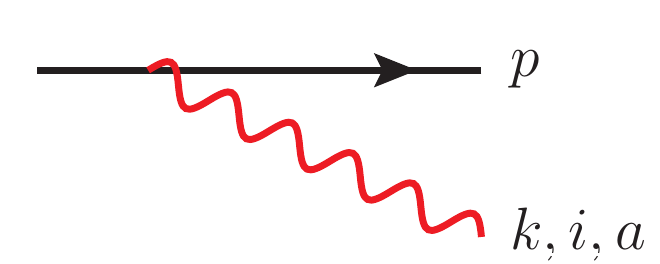} \end{minipage} &= \hspace{0.5em} \VG^{a,i}(\kap,z) \,,\\
\begin{minipage}{4.5cm} \includegraphics[width=\textwidth]{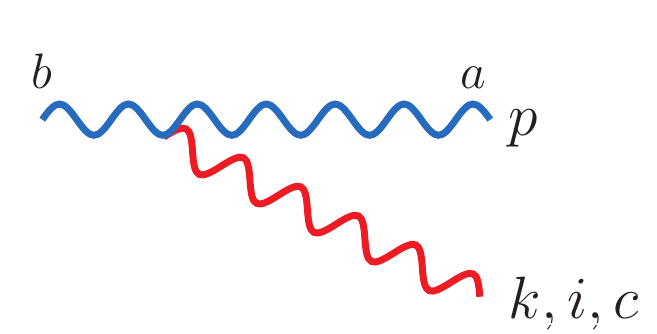} \end{minipage} &= \hspace{0.5em} \GamG^{abc,i}(\kap,z) \,,
\end{align}
\begin{align}
\begin{minipage}{4.3cm} \includegraphics[width=\textwidth]{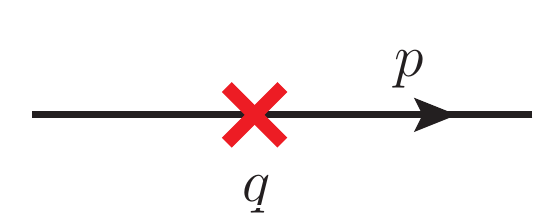} \end{minipage} \hspace{1em}&= \hspace{0.5em} \VA(x^+;p^+,\q) \,,\\
\begin{minipage}{4.5cm} \includegraphics[width=\textwidth]{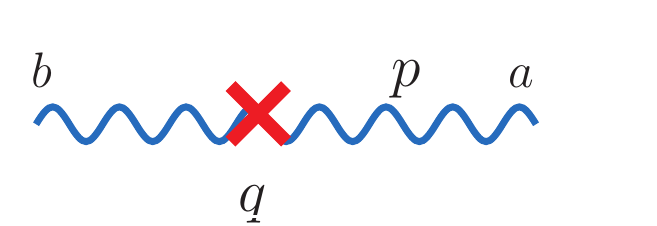} \end{minipage} &= \hspace{0.5em} \GamA^{ab}(x^+;p^+,\q) \,.
\end{align}
Finally, we will discuss the situation where the same parton line interacts twice with the medium. Due to the instantaneous nature of the interactions, see \eq{eq:MediumAverage}, the scalar propagator in between the two medium-insertions reduces to $\Theta(0)/(2k^+) = 1/(4k^+)$ and the double-interaction is denoted by a circle instead of two crosses. Since we can perform the integrations over the medium-momentum explicitly, we obtain the following two rules
\begin{align}
\begin{minipage}{4.3cm} \includegraphics[width=\textwidth]{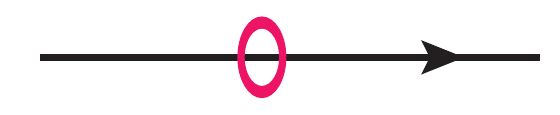} \end{minipage} \hspace{1em}&= \hspace{0.5em} -\frac{1}{2}\alpha_s C_F \int \dd x^+\,n(x^+) \,,\\
\begin{minipage}{4.3cm} \includegraphics[width=\textwidth]{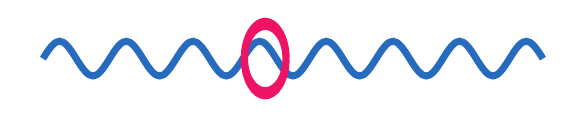} \end{minipage}\hspace{1em} &= \hspace{0.5em} \frac{1}{2}\alpha_s C_A \int \dd x^+\,n(x^+) \,,
\end{align}
where, in order to obtain explicit expressions, we have assumed that $\Vscat = (\q^2 +\mDebye^2)^{-2}$. However, in our Mathematica code, we have calculated these diagrams using the automated procedure similarly to all the other diagrams.

Additionally, all newly produced final-state partons, i.e., partons that propagate from some vertex to the cut, have to be multiplied by the appropriate polarization vector, for gluons, or spinor, for fermions, and by the appropriate phase $e^{i p^-x^+}$, where $p$ is the parton momentum and $x^+$ the position of the last vertex. We summarise these rules below:
\begin{align}
\begin{minipage}{3.5cm} \includegraphics[width=\textwidth]{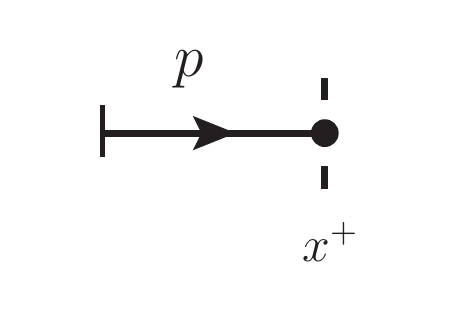} \end{minipage} &= e^{-ip^- x^+} u^s(p)  \,,\\
\begin{minipage}{3.5cm} \includegraphics[width=\textwidth]{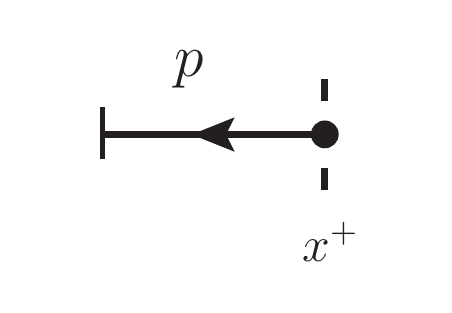} \end{minipage} &= e^{-ip^- x^+} \bar u^s(p)  \,,\\
\begin{minipage}{3.5cm} \includegraphics[width=\textwidth]{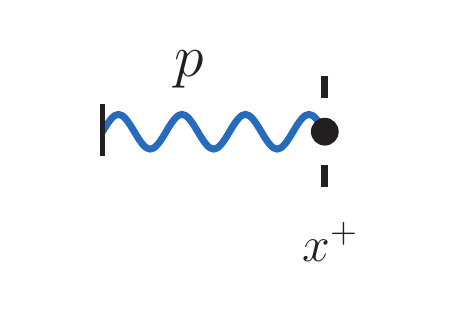} \end{minipage} &= e^{-ip^- x^+} \epsk_\lambda^i (p)  \,,
\end{align}
and similarly for final-state particles,
\begin{align}
\begin{minipage}{3.5cm} \includegraphics[width=\textwidth]{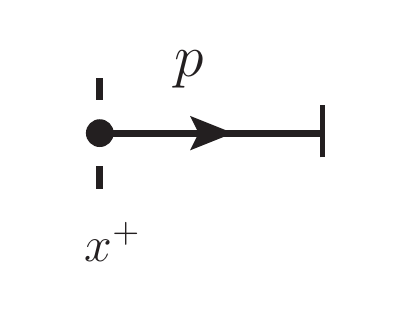} \end{minipage} &= e^{ip^- x^+} \bar u^s(p)  \,,\\
\begin{minipage}{3.5cm} \includegraphics[width=\textwidth]{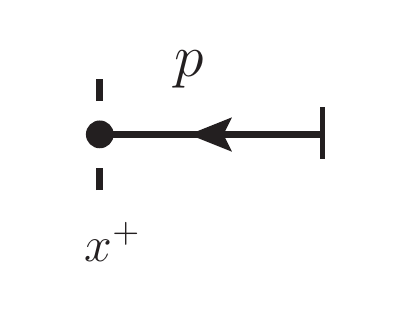} \end{minipage} &= e^{ip^- x^+} u^s(p)  \,,\\
\begin{minipage}{3.5cm} \includegraphics[width=\textwidth]{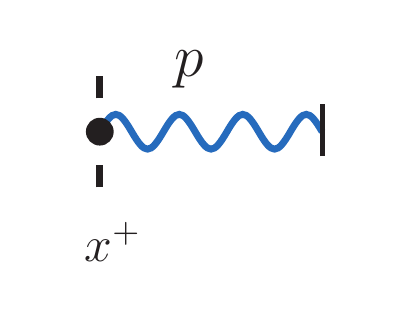} \end{minipage} &= e^{ip^- x^+} \epsk^{\ast\, i}_\lambda(p)  \,,
\end{align}
where the ``cut'' (corresponding to $x^+ = \pm \infty$) is represented by a small vertical line.




\section{Short derivation of the antenna spectrum in medium}
\label{sec:AntennaDerivation}

Since soft gluon radiation can be treated as a classical process, we can apply methods from classical Yang-Mills theory to obtain the amplitude for an emission off one of the legs of the antenna, which reads
\begin{align}
\label{eq:AntenaAmplitude}
\mathcal{M}_{\lambda,1}^a &= 2ig \intq \int_0^\infty\dd x^+ \left[T\cdot \mathcal{A}(x^+;\q) \right]^{ab} \Q_1^b \nn
 &\times \left\{ \Aone - \left[\Aone - \Bone\right]e^{i\frac{(\kapone-\q)^2}{2 k^+}x^+} \right\} \cdot \epsk_\lambda(\k) \,,
\end{align}
where
\beq
\kapone = \k - \frac{k^+}{p_1^+}\pone \,,
\eeq
is the transverse momentum of the gluon $\k$ with respect to the transverse momentum of one of the legs of the antenna and $[T\cdot \mathcal{A}]^{ab}\Q^b_1 = i f^{abc} \mathcal{A}^b \Q^c_1$. The full amplitude, corresponding to emissions of both legs reads simply $\mathcal{M}_{\lambda,1+2}^a = \mathcal{M}_{\lambda,1}^a+\mathcal{M}_{\lambda,2}^a$. The inclusive one-gluon cross section reads then
\beq
\frac{\dd N}{\dd^3k} = \frac{1}{(2\pi^3)2 k^+} \sum_{\lambda,\lambda', a, a'} \langle \mathcal{M}_{\lambda,1+2}^a \mathcal{M}_{\lambda',1+2}^{\ast, a'} \rangle \,,
\eeq
where the brackets $\langle \ldots \rangle$ imply the medium average defined in \eq{eq:MediumAverage}. The medium-induced spectrum is obtained from this after redifining the potential as
\beq
\Vscat \to \Vscat - (2\pi)^2 \delta (\q) \int_{{\bs q'}} \mathcal{V}(\q') \,,
\eeq
in order to account for virtual corrections in a completely analogous manner as the medium-virtual diagrams considered in the Feynman diagram technique utilized in the remainder of the paper.

A very similar situation to the one we are considering in the current work, is the emission of a soft gluon off a colour dipole, or usually called an ``antenna''. We will label each of the emitters in this case simply by ``1'' and ``2'', and their kinematics is given by $p_i=(p_i^+,p^-_i,\p_i)$ while the momentum of the emitted gluon is $k=(k^+,k^-,\k)$. Using the results in \cite{MehtarTani:2011gf}, we will here generalise their results for an antenna in general colour configuration. 

In vacuum, the square of the emission amplitude, summed over colours and polarizations, reads,
\beq
\label{eq:Ampl2Vacuum}
\mathcal{P}^{(0)}_\text{ant}(k) \equiv \sum_{\lambda} \left| \mathcal{M}_\lambda^{(0)}\right|^2 = 4g^2 \,\left(\Q_1^2\frac{1}{\kapone^2} + \Q_2^2 \frac{1}{\kaptwo^2} + 2 \Q_1\cdot \Q_2 \frac{\kapone\cdot \kaptwo}{\kapone^2 \kaptwo^2} \right) \,,
\eeq
where 
\beq
\label{eq:kappaDef}
{\bs \kappa}_i \equiv \k - z_i {\bs p}_i \,,
\eeq
with $z_i \equiv k^+/p^+_i$ is the light-cone momentum fraction, is the transverse momentum of the emitted gluon with respect to the emitting antenna constituent.

In order to simplify the colour algebra, we have introduced the colour vectors $\QC^a_1$ and $\QC^a_2$ that obey the following property $\QC^a_1+\QC^a_2 = \QC^a_3$, where $\QC^2_3 \equiv \QC^a_3 \cdot \QC^a_3$ is the total charge of the antenna. These vectors are defined such that for a quark $\QC^2_q = C_F$ while for a gluon $\QC^2_g = C_A$, where $C_F = (N_c^2-1)/(2N_c)$ and $C_A = N_c$. Finally, by squaring this relation we solve for the cross-term to find $\QC_1\cdot \QC_2 = (\QC^2_3 - \QC_1^2 - \QC^2_2)/2$. The possible QCD $1\to 2$ splittings give
\begin{align}
\label{eq:AntennaColor}
\QC_1^2=\QC_2^2 = C_F \text{,   and } \QC^2_3 = C_A & \hspace{1em}\text{ for } g \to q+\bar q \,,\\
\QC_1^2=\QC_2^2 = \QC_3^2 = C_A & \hspace{1em}\text{ for } g \to g+g \,,\\
\label{eq:qqgColor}
\QC_1^2=\QC_3^2 = C_F \text{,   and } \QC^2_2 = C_A & \hspace{1em}\text{ for } q \to q+g \,.
\end{align}
While the first two  situations were analysed in references \cite{MehtarTani:2010ma,MehtarTani:2011tz,Armesto:2011ir,CasalderreySolana:2011rz,MehtarTani:2011gf,MehtarTani:2012cy}, 
we are mostly interested in the latter process for the moment. Let us also introduce a compact notation that will prove very useful in the following sections. First, defining the building blocks
\begin{align}
\Avec^a_1 &\equiv \QC^a_1 \Aone  \,,\\
\Bvec^a_1 &\equiv \QC^a_1 \Bone \,,
\end{align}
where the former will come in handy for the medium part. Note that $\Avec^a$ and $\Bvec^a$ are both transverse vectors and vectors in colour space, dropping the superscript ``$a$'' simply defines the corresponding transverse vector. The same goes for any other similarly defined vector below.

For the colour configuration (\ref{eq:qqgColor}), the vacuum emission antenna spectrum takes the form
\beq
\label{eq:Vacuumqqg}
\mathcal{P}^{(0)}_\text{ant}(k) = 4g^2 \left[ C_F \Bvec_1^2 + C_A \left(\Bvec_2^2 - \Bvec_1\cdot \Bvec_2 \right) \right] \,.
\eeq
Due to the colour algebra, the antenna quark radiates as a free one. This comes about due to the combination of the radiation inside the cone off the antenna legs and the large-angle radiation outside the cone by the total charge. Additionally, the gluonic antenna leg can radiate inside the cone.

Now proceeding to the situation where one medium interaction is allowed, we define the currents
\begin{align}
\label{eq:LipatovVertex}
\Lip^a_1 \equiv \Avec^a_1 - \Bvec^a_1 \,, \\
\Cvec^a \equiv \Avec^a_1 + \Avec^a_2 \,,
\end{align}
where the former, \eq{eq:LipatovVertex}, often is referred to as the ``Lipatov vertex''. Using this notation, it is possible to write the squared amplitude, after taking the medium average and summing over spins and colours, $\Mantenna(k) \equiv \sum_\lambda \langle |\mathcal{M}^{(1)}_\lambda |^2\rangle$, as
\beq
\label{eq:Ampl2Medium_1}
\Mantenna(k) = 4 g^4 C_A \, \mDebye^2 \, \intq \mathcal{V}^2(\q) \int_0^\infty \dd x^+ \, n\left(x^+\right) \tilde w^{(1)}_\text{ant}\left(x^+; k,\q\right) \,,
\eeq
where
\begin{align}
\label{eq:Ampl2Medium_2}
\tilde w^{(1)}_\text{ant}\left(x^+; k,\q\right) &= 2\left[1- \cos \frac{x^+}{\tone} \right] \Lip_1^a\cdot \Cvec^a + 2\left[1-\cos\frac{x^+}{\ttwo} \right] \Lip_2^a\cdot\Cvec^a \nn
& - 2\left[1-\cos\frac{x^+}{\tonetwo} \right] \Lip_1^a\cdot \Lip_2^a \,,
\end{align}
and $\tone = 2 k^+\big/\left(\kapone - \q \right)^2$, $\ttwo = 2 k^+\big/\left(\kaptwo - \q \right)^2$ and $\tonetwo = \big(1/\tone - 1/\ttwo \big)^{-1}$. While $\tone$ and $\ttwo$ are simply the formation times of a medium-induced gluon off either of the legs of the antenne, $\tonetwo$ sets the time-scale for interference effects. Performing the colour decomposition, as given by \eq{eq:qqgColor}, allows us to write 
\begin{align}
\label{eq:Ampl2Medium_Color}
\tilde w^{(1)}_\text{ant}\left(x^+; k,\q\right) &= C_F \left[1- \cos \frac{x^+}{\tone} \right] \left(\Lip_1^2 + \Avec_1^2 -\Bvec_1^2 \right) \nn
& + C_A\Bigg\{\left[1- \cos \frac{x^+}{\ttwo} \right] \left(\Lip_2^2 + \Avec_2^2 -\Bvec_2^2 - \Avec_1\cdot\Lip_2 \right) \nn
&\hspace{1em} - \left[1- \cos \frac{x^+}{\tone} \right]\, \Avec_2\cdot\Lip_1 +\left[1- \cos \frac{x^+}{\tonetwo} \right]\,  \Lip_1\cdot \Lip_2 \Bigg\} \,.
\end{align}
In order to make some sense out of this complicated expression, let us take the completely coherent scattering limit, i.e. $x^+ \to \infty$. In that case we can neglect all the cosines, and \eq{eq:Ampl2Medium_Color} reduces to
\begin{align}
\label{eq:Ampl2Medium_Coherent}
\lim_{x^+ \to \infty} \tilde w^{(1)}_\text{ant}\left(x^+; k,\q\right) &= C_F \left[\Lip_1^2 + \Avec_1^2 -\Bvec_1^2 \right] \nn
& + C_A \left[ \Lip_2^2 + \left(\Avec_2^2 - \Avec_1\cdot \Avec_2 \right) - \left( \Bvec_2^2 - \Bvec_1\cdot \Bvec_2\right) \right] \,.
\end{align}
We see here clearly the expected features. In the medium, the Lipatov vertex gives rise to the Gunion-Bertsch spectrum, represented by $\Lip_i^2$. Secondly, the pure vacuum spectrum is affected by broadening, effectively replacing the vacuum spectrum with a broadened one with the appropriate weight. For the emission off the quark, since all interferences cancel, this is represented by $\Avec_1^2 - \Bvec^2_1$. For the emission off the gluon, due to the presence of the interference in the term proportional to $C_A$ in \eq{eq:Vacuumqqg}, we replace the complete vacuum contribution completely analogously. A similar systematic was found for the $q\bar q$ antenna in \cite{MehtarTani:2011gf}.

The analysis of the antenna spectrum for $q\to q+g$ has thus provided us with crucial information. Whenever the medium scattering takes place long after the gluon formation and interference times, we see simply that $\W^{(1)}$ subtracts a piece of the vacuum radition from $\W^{(0)}$, with a proper weight, see \eq{eq:Ampl2Medium_1}, and replaces it with a broadening. In addition to this, all colour charges radiate a Gunion-Bertsch spectrum. Naturally, in the general case, for finite scattering times, we obtain a quite complicated interference pattern, \eq{eq:Ampl2Medium_Color}. This can however be used to compare the full $q\to q+g$ splitting with the antenna case, which represents the case when the formation time of the parent quark is exactly zero.

\clearpage

\bibliography{TwoGluon}
\bibliographystyle{JHEP}

%

\end{document}